\documentclass[aps,prd,twocolumn,superscriptaddress,preprintnumbers,floatfix,nofootinbib,longbibliography,showlabels]{revtex4-1}

\usepackage[utf8x]{inputenc}
\usepackage{float}
\usepackage{graphicx}
\usepackage{color} 
\usepackage{hyperref}
\usepackage{epsfig}
\usepackage{wrapfig}
\usepackage{amsmath}
\usepackage{amssymb}
\usepackage{hyperref}
\usepackage{soul}

\def\lsim{\raise0.3ex\hbox{$<$\kern-0.75em\raise-1.1ex\hbox{$\sim$}}}
\def\gsim{\raise0.3ex\hbox{$>$\kern-0.75em\raise-1.1ex\hbox{$\sim$}}}

\newcommand{\tr}{\textrm{Tr}\,}

\newcommand{\beq}{\begin{eqnarray}}
\newcommand{\eeq}{\end{eqnarray}}
\newcommand{\be}{\begin{equation}}
\newcommand{\ee}{\end{equation}}

\newcommand{\bmat}{\left (\begin{array}{cc}}
\newcommand{\emat}{\end{array} \right )}

\definecolor{grey}{RGB}{100,100,100}

\newcommand{\half}{\mbox{${\frac12}$}}

\newcommand{\pd}{\partial}
\newcommand{\rmd}{\mathrm{d}}
\newcommand{\A}{\alpha}
\newcommand{\B}{\beta}
\newcommand{\C}{\gamma}
\newcommand{\D}{\delta}

\begin{document}

\preprint{HIP-2025-19/TH}
 
\title{Non-perturbative determination of the sphaleron rate for first-order phase transitions}

\author{Jaakko Annala}
\email{jaakko.annala@helsinki.fi}
\affiliation{Department of Physics and Helsinki Institute of Physics, PL 64 (Gustaf H\"allstr\"omin katu 2), FI-00014 University of Helsinki, Finland}

\author{Kari Rummukainen}
\email{kari.rummukainen@helsinki.fi}
\affiliation{Department of Physics and Helsinki Institute of Physics, PL 64 (Gustaf H\"allstr\"omin katu 2), FI-00014 University of Helsinki, Finland}

\author{Tuomas V.I. Tenkanen}
\email{tuomas.tenkanen@helsinki.fi}
\affiliation{Department of Physics and Helsinki Institute of Physics, PL 64 (Gustaf H\"allstr\"omin katu 2), FI-00014 University of Helsinki, Finland}

\begin{abstract}

In many extensions of the Standard Model electroweak phase transitions at high temperatures can be described in a minimal dimensionally reduced effective theory with SU(2) gauge field and fundamental Higgs scalar.  
In this effective theory, all thermodynamic information is governed by two dimensionless ratios
$x \equiv \lambda_3/g^2_3$ and $y\equiv m^2_3/g^4_3$, where $\lambda_3$, $m^2_3$ and $g_3$ are the effective thermal scalar self-interaction coupling, the thermal mass and the  effective gauge-coupling, respectively.
By using non-perturbative lattice simulations to determine the rate of sphaleron transitions in the entire $(x,y)$-plane corresponding to the Higgs phase,
and by applying previous lattice results for the bubble nucleation,
we find a condition $x(T_c) \lesssim 0.025$ to guarantee preservation of the baryon asymmetry, which translates to $v/T_c \equiv \sqrt{2 \Delta \langle \phi^\dagger \phi \rangle}/T_c \gtrsim 1.33$ for the (gauge-invariant) discontinuity in Higgs condensate. This indicates that viability of the electroweak baryogenesis requires the phase transition to be slightly stronger than previously anticipated.
Finally, we present a general template for analysing such viability in a wide class of beyond the Standard Model theories, in which new fields are heavy enough to be integrated out at high temperature. 
\end{abstract}

\maketitle

\section{Introduction}
\label{sec:intro}

Explanation to the observed matter/antimatter asymmetry of the universe still remains elusive, despite long-standing efforts \cite{Bodeker:2020ghk}. Among the most studied candidates for a dynamical generation of the asymmetry is the electroweak baryogenesis \cite{Rubakov:1996vz,Morrissey:2012db}, where baryonic excess over their antimatter counterparts is generated during an electroweak phase transition. Such a phase transition is a violent process, where macroscopic bubbles of the low-temperature Higgs phase nucleate and eventually fill the young cosmos. Since the detection of gravitational waves from astrophysical sources a decade ago \cite{LIGOScientific:2016aoc}, attempts to understand such phase transitions have become fashionable, due to their potential to source a stochastic background of gravitational waves \cite{Caprini:2015zlo,Caprini:2019egz,LISACosmologyWorkingGroup:2022jok}. Detecting such a background is among the objectives of the future LISA mission \cite{LISA:2017pwj} and other next-generation space-borne gravitational wave experiments \cite{Kawamura:2006up,Harry:2006fi,Ruan:2018tsw}.

While the Standard Model of particle physics (SM) with the minimal scalar sector lacks a first-order electroweak phase transition \cite{Kajantie:1996mn,Aoki:1996cu,Gurtler:1997hr,Csikor:1998eu,Rummukainen:1998as}, in many of its extensions a first-order transition can be triggered 
\cite{Cline:1996cr,Laine:1996ms,Laine:1998qk,Laine:1998vn,Laine:2000rm,Laine:2012jy,Kainulainen:2019kyp,Niemi:2020hto,Niemi:2024axp}. 
This requires that new scalar fields, relevant at the electroweak scale, are light enough and are not too weakly interacting with the Higgs field, making such models attractive targets for future collider experiments \cite{Ramsey-Musolf:2019lsf}. 

As a result of the non-trivial topological vacuum structure of SU(2) gauge theory
(in which classically equivalent but topologically distinct vacua are enumerated by a topological Chern-Simons number \cite{Belavin:1975fg})
baryon number can be violated through the process of Chern-Simons number diffusion.
In the high-temperature confinement phase where the Higgs mechanism is not active this is a purely non-perturbative process. Once the thermal plasma is devoured by the expanding bubble of the low-temperature Higgs phase the activation of the Higgs mechanism allows us to understand baryon violating processes in terms of sphalerons. 
In a perturbative picture (applicable in the Higgs phase), sphalerons are semi-classical unstable saddle-point configurations of the Higgs and SU(2) gauge fields \cite{Klinkhamer:1984di}. 
Sphaleron configurations correspond to half-integer Chern-Simons number that lie on top of an energy barrier separating two topologically distinct vacua with consecutive integer Chern-Simons numbers. Thermal fluctuations across the sphaleron barrier can violate the baryon number through the (Adler-Bell-Jackiw) chiral anomaly \cite{Adler:1969gk,Bell:1969ts,tHooft:1976snw}.

The sphaleron transitions (or Chern-Simons number diffusion) are very rapid in the confinement phase, yet exponentially suppressed in the Higgs phase.
Consequently, if the expanding bubble wall introduces enough charge (C) and charge-parity (CP) violation in front of the bubble wall, sphalerons in the confinement phase can turn such bias into a net baryon number \cite{Kuzmin:1985mm,PhysRevD.36.581}. The excess of baryons is then absorbed by the expanding bubble, and if the rate of sphaleron transitions is sharply shut off in the Higgs phase, 
the baryon number freezes to its value observed today, providing us a mechanism for baryogenesis.

While such a process has multiple stages, in this article our goal is to tackle only one of them: to accurately determine the sphaleron rate in scenarios that have a first-order electroweak phase transition. To achieve this, we resort to non-perturbative lattice simulations in a minimal dimensionally reduced effective theory for the electroweak phase transition. Such a minimal setup is provided by the theory with SU(2) gauge field and fundamental Higgs scalar, for which the ratio of thermal effective Higgs self-interaction coupling and the effective gauge-coupling $x \equiv \lambda_3/g^2_3$
at the critical temperature controls the character of phase transition, first-order transitions existing for $x<x_* = 0.0983$ \cite{Kajantie:1995kf,Gould:2022ran}. This condition can be fulfilled in a class of theories where new scalars are sufficiently heavy at high temperatures (but not necessarily at zero temperature) that they can be integrated out from the thermal effective theory, while reducing the effective Higgs self-interaction coupling and thereby rendering the phase transition to be of first-order. Typically, this requires a large portal interaction with the Higgs field, enhancing detection prospects with future particle colliders.

Our present study combines the previous non-perturbative knowledge of thermodynamics and bubble nucleation rate \cite{Moore:2000jw,Gould:2022ran} within the minimal framework while adding the non-perturbatively determined condition for the sphaleron freeze-out in terms of dimensionless ratios $x$ and $y$. The fact that a full class of possible beyond the Standard Model (BSM) scenarios can be mapped into the minimal thermal effective theory, makes it possible to recast this non-perturbative information to any such parent theory, merely in terms of a dimensional reduction mapping into the effective field theory; a procedure which by now is to a large extend fully automated \cite{Ekstedt:2022bff} (see e.g.~\cite{Tenkanen:2022tly,Curtin:2022ovx,Kierkla:2023von,Ekstedt:2024etx,Athron:2024xrh,Karkout:2024ojx,Bertenstam:2025jvd,Brdar:2025gyo}).

Furthermore, as a by-product, we also provide a comprehensive comparison to a recent fully perturbative approach tackling the same problem  \cite{Li:2025kyo}.
Such comparison has also further-reaching motives: compared to perturbation theory, non-perturbative lattice studies are expensive, taking considerable computation time. At present, such simulations for the sphaleron rate are well established for the minimal effective theory we work with \cite{Moore:1998swa,Moore:2000mx,DOnofrio:2012phz,DOnofrio:2014rug,Annala:2023jvr}, as well as for the bubble nucleation rate \cite{Moore:2000jw,Gould:2022ran} (see also \cite{Moore:2001vf,Gould:2024chm,Hallfors:2025key}). Yet, for more complicated models, which include new fields within the effective theory, similar non-perturbative studies have not yet been performed, albeit non-perturbative simulations of equilibrium properties have appeared in \cite{Laine:1998qk,Laine:1998vn,Laine:2000rm,Laine:2012jy,Kainulainen:2019kyp,Niemi:2020hto,Niemi:2024axp}. However, establishing a reliable perturbative approach -- i.e. concretely learning which perturbative effects have to be included and what are their limitations -- makes it possible to tackle these more complicated models by purely perturbative approach, which would allow a scan through their often vast parameter spaces with relative ease.

Rest of this article is organised as follows:
In Sec.~\ref{sec:simu} we review the setup for the simulations which aligns with those in \cite{Annala:2023jvr}.
In sections \ref{sec:results-lattice} and \ref{sec:results-cosmo} we present our results and compare them to a recent purely perturbative study \cite{Li:2025kyo}, while relegating further discussion on perturbative descriptions to Appendix~\ref{sec:pert}.
Sec.~\ref{sec:BSM} is devoted to a discussion of the class of parent theories that admit a first-order transitions and can be mapped into the minimal effective theory.
A summary and discussion of our findings is presented in Sec.~\ref{sec:conc}. 
In Appendix~\ref{sec:temporal} we briefly discuss the validity of integrating out the temporal gauge field modes for strong transitions.

\section{Setup for the simulations}
\label{sec:simu}

The sphaleron rate%
\footnote{
The term sphaleron rate is often used loosely to refer to the Chern-Simons number diffusion constant in both phases, while technically the term sphaleron coined in \cite{Klinkhamer:1984di} refers to the classical saddle-point configuration, which only exists in the Higgs phase.   
Also, one often encounters alternative terminology for the two phases ``symmetric/broken'', yet we avoid this as local gauge symmetry cannot be spontaneously broken \cite{Elitzur:1975im}. 
}
is the diffusion constant for the Chern-Simons number $N_{\rm CS}$, 
which is defined through the SU(2) field strength $F_{\alpha\beta}$ as \cite{Kuzmin:1985mm,Rubakov:1996vz,Moore:2010jd} 
\begin{align}
N_{\rm CS}(t)-N_{\rm CS}(0)&\equiv \frac{g^2}{32\pi^2}\int_0^t \rmd t \int \rmd^3x \epsilon_{\A\B\C\D} \tr F^{\A\B} F^{\C\D} \ ,
\end{align}
where Greek letters are Lorentz indices and $\epsilon_{\A\B\C\D}$ is the totally antisymmetric tensor with $\epsilon_{0123}=+1$.

Akin to \cite{Farakos:1994xh,DOnofrio:2012phz,DOnofrio:2014rug,Annala:2023jvr} our starting point for the simulations is a dimensionally reduced effective field theory (EFT) in three spatial dimensions (3D) \cite{Ginsparg:1980ef,Appelquist:1981vg,Braaten:1995cm,generic}, defined through the following 
continuum action%
\footnote{In this EFT, temporal gauge field components have been integrated out \cite{generic}. For validity of this, see Appendix~\ref{sec:temporal}. }
\begin{align}
\label{eq:continuum_theory}
L_{\mathrm{3D}} = \frac{1}{4} \tr F_{ij} F_{ij} + (D_i\phi)^\dagger (D_i\phi) +m_3^2\phi^\dagger\phi + \lambda_3(\phi^\dagger\phi)^2, 
\end{align}
where the field strength and covariant derivative are
\begin{align}
    F_{ij} &= \pd_iA_j -\pd_jA_i - g_3 [A_i, A_j]\ , ~~~ A_i = \half\sigma_b A_i^b \ , \nonumber\\
    D_i    &= \pd_i+ig_3A_i \ ,
\end{align}
for the 3D SU(2) gauge field $A_i$, with the dimensionful SU(2) coupling $g_3$. 
Indices $i,j=1,2,3$ denote spatial Lorentz indices, while $b=1,2,3$ is the SU(2) adjoint index.
Finally, $\phi$ is a complex scalar doublet, for which we define dimensionless parameters
\begin{equation}
  x(T)\equiv \frac{\lambda_3}{g_3^2},\qquad
  y(T)\equiv \frac{m_3^2}{g_3^4},
  \label{3d_params}
\end{equation}
via thermal scalar self-interaction and thermal mass, respectively. We do not include U(1) gauge field (i.e. we set the Weinberg angle $\theta_W=0$) within the EFT, since it has a negligible effect on the phase transition \cite{Kajantie:1996mn,Kajantie:1996qd,DOnofrio:2015gop,Laine:2015kra} and the sphaleron rate%
\footnote{
We assume here that there are no strong external hypermagnetic fields. Strong hypermagnetic fields affect both the phase transition and the sphaleron rate, see e.g. refs.~\cite{Annala:2023jvr,Kajantie:1998rz} for non-perturbative studies.
}
\cite{Klinkhamer:1990fi, Kleihaus:1991ks, Kunz:1992uh,Annala:2023jvr}. Eq.~\eqref{eq:continuum_theory} is the effective theory of the SM, but also of many of its extensions, provided that new BSM fields are heavy enough (akin to temporal gauge field components) at and below the critical temperature of the transition. We provide concrete examples of this in Sec.~\ref{sec:BSM}.

Phase diagram of this EFT is dictated by the curve $y(T_c) = y_c(x)$, which has been determined non-perturbatively in \cite{Kajantie:1995kf,Gould:2022ran} (as well as perturbatively in \cite{Kajantie:1995kf,Ekstedt:2022zro, Ekstedt:2024etx} to maximal perturbatively available order \cite{Linde:1980ts}). Below the critical $y_c(x)$ curve (after supercooling to temperatures below the critical temperature), one can associate a similar curve $y_p(x)$ for the condition that the system percolates, and the transition completes \cite{Gould:2022ran}.
This curve can be determined from combining a condition from cosmological evolution with the bubble nucleation rate, which has been measured non-perturbatively for four different values of $x$ in \cite{Gould:2022ran}, as well as calculated perturbatively in \cite{Baacke:1999sc,Ekstedt:2021kyx,Ekstedt:2022ceo,Gould:2022ran}. 

Here, we find non-perturbatively an analogous curve $y_f(x)$ for a condition of the sphaleron freeze-out in the Higgs phase \cite{Burnier:2005hp}. Recently, such a curve was determined in a semi-classical approximation within the EFT in \cite{Li:2025kyo}. As formulated in \cite{Li:2025kyo}, if the sphaleron freeze-out occurs at temperatures higher than the percolation temperature, any baryon number washout is prevented inside the bubble, and hence any excess generated at the bubble wall would automatically be preserved inside the bubble, potentially leading to successful baryogenesis.%
\footnote{
We note, that this conclusion relies on an assumption that the system remains in thermal equilibrium, yet across the bubble wall the system departs from equilibrium and temperature varies across the bubble. For deflagrations, the bubble can heat the plasma in the Higgs phase inside the bubble back towards the critical temperature, c.f~\cite{Enqvist:1991xw}. We have implicitly assumed that such out-of-equilibrium effects are negligible.
}

The lattice action corresponding to the dimensionally reduced 3D theory \eqref{eq:continuum_theory} is given by
\begin{align}\label{lattice_action}
    S&= \beta_G \sum_x \sum_{i<j}[1-\half \tr P_{ij}(x)] \nonumber\\
     & - \beta_H \sum_x \sum_{i} \half\tr\Phi^\dagger(x)U_i(x)\Phi(x+\hat i) \\
     & + \beta_2 \sum_x \half \tr\Phi^\dagger(x)\Phi(x) + \beta_4\sum_x
         \big[\half\tr\Phi^\dagger(x)\Phi(x)\big]^2, \nonumber
\end{align}
where $U_i(x)$ are the SU(2) link variables, $P_{ij}(x)$ are the SU(2) plaquettes defined as
\begin{align}
      P_{ij}(x) &= U_i(x)U_j(x+\hat i)
      U^\dagger_i(x+\hat j)U^\dagger_j(x) \ , \label{su2plaq}
\end{align}
and the Higgs field is written in matrix form
\begin{equation}
    \Phi \equiv \frac{1}{g_3^2} 
    \left(\begin{array}{cc}
        \phi_2^*&\phi_1\\
        -\phi_1^*&\phi_2
      \end{array}\right) \ .
    \label{phimatrix2}
\end{equation}
The lattice parameter setting the scale of the lattice is $\beta_G = 4/(g_3^2 a) + 0.6674...$ where $a$ is the lattice spacing. The numerical factor comes from $O(a)$ improvements to the lattice continuum relations where the renormalisation scale is chosen as $\mu_3 = g_3^2$. The rest of the lattice parameters $\beta_Y, \beta_H, \beta_2, \beta_4$ are similarly related to the continuum parameters $g_3^2$, $x$ and $y$ through functions computable in lattice perturbation theory \cite{Laine:1995np,Laine:1997dy}. The partial $O(a)$ improvements to the lattice parameters we use can be found in \cite{Moore:1996bf,Moore:1997np}. 
The relations, with the same notation as this paper, can be found in Appendix A in Ref.~\cite{Annala:2023jvr} (setting the U(1) coupling to zero $z\to 0$ which is not included in this work as described above).

Following previous works \cite{DOnofrio:2012phz,DOnofrio:2014rug,Annala:2023jvr} we use the effective Langevin type description for the dynamical evolution. The dynamics of the soft modes that govern the evolution of the sphaleron configurations is fully overdamped and given by an effective Langevin type evolution to a leading logarithmic accuracy $1/\ln(1/g)$ \cite{Bodeker:1998hm,Arnold:1998cy,Bodeker:1999ey,Arnold:1999jf,Arnold:1999uy} in terms of SU(2) coupling $g$ of the parent theory.

In the Higgs phase the sphaleron rate can be computed from the lattice as \cite{Moore:1998swa,Moore:2000jw}%
\footnote{
Technically, this is the Chern-Simons number diffusion rate, which is twice the sphaleron rate \cite{Burnier:2005hp}. However, we refer Eq.~\eqref{sph_rate_lattice} as the sphaleron rate in order to stay consistent with terminology used in \cite{Annala:2023jvr}.
}
\begin{equation}
\label{sph_rate_lattice}
    \tilde{\Gamma}(x,y) = \underbrace{\frac{P(|N_{\rm CS}-\tfrac{1}{2}|<\tfrac{\epsilon}{2})}{\epsilon V}}_{\text{statistical}} \underbrace{\left\langle \left| \frac{\Delta N_{\rm CS}}{\Delta t} \right| \right\rangle \rmd }_{\text{dynamical}}\ ,
  \end{equation}
taking here the lattice volume $V$ and the small time-step $\Delta t$ to be in dimensionless lattice units. The rate factorises to statistical, equilibrium and dynamical, out-of-equilibrium parts. 

The statistical part $P(N_{\rm CS})$ is obtained from multicanonical Monte Carlo computation which determines the probabilistic suppression of getting to a sphaleron configuration at $N_{\rm CS} = \tfrac{1}{2} \pm \tfrac{\epsilon}{2}$ with some small tolerance $\epsilon \ll 1$ (we have used $\epsilon = 0.04$). While the dynamical part determines the rate of transitioning from one vacuum to another at the top of barrier at, or close to, the sphaleron configuration. This is computed from evolving configurations at $N_{\rm CS} = \tfrac{1}{2} \pm \tfrac{\epsilon}{2}$ in time and measuring the number of trajectories that evolve from one vacuum to another. 
A single trajectory may cross the sphaleron barrier multiple times and the ``dynamical prefactor'' $\rmd$ is computed to account for this effect. The prefactor is defined as the sum over the simulated sample of trajectories $\rmd = \sum_{traj} \delta_{\rm tunnel}/(N_{\rm cross}N_{\rm traj})$ where $\delta_{\rm tunnel} = 1$ if the trajectory changes the vacuum $\Delta N_{\rm CS} = \pm 1$ and $\delta_{\rm tunnel} = 0$ otherwise, and $N_{\rm cross}$ is the number of times the trajectory crosses the sphaleron barrier at $N_{\rm CS}=\tfrac{1}{2}$. See \cite{Moore:1998swa,Moore:2000jw,DOnofrio:2012phz} for more details.

The naive lattice discretisation of the topological Chern-Simons number contains ultraviolet noise leading to unphysical diffusion which can be removed by cooling the configuration towards vacuum by a set amount of fictitious cooling time, which gets rid of the unphysical noise.
Due to the cooling used in the measurement, the statistical and dynamical parts of the rate \eqref{sph_rate_lattice} do not separately have a simple physical interpretation. Cooling the configuration brings it closer to the vacuum decreasing the probability of finding a configuration with half integer Chern-Simons number, thus decreasing $P_\epsilon \equiv P(|N_{\rm CS}-\tfrac{1}{2}|<\tfrac{\epsilon}{2})$. On the other hand, cooling will similarly increase $\langle \Delta N_{\rm CS} / \Delta t \rangle$ by the same factor as it decreases $P_\epsilon$ keeping the product of the two independent of the cooling time.

Physical rate relates to the dimensionless quantity $\tilde{\Gamma}$ computed from the lattice as
\begin{equation}
    \Gamma(x,y) = \frac{(g_3^2)^5}{\sigma_{\textrm{el}}} \tilde{\Gamma}(x,y),
\end{equation}
where $\sigma_{\textrm{el}}$ is the SU(2) color conductivity (that describes
the current response to external fields at the infrared) 
computed via hard thermal loop effective description, which is given as \cite{Arnold:1999uy}:
\begin{align}
\label{eq:color-conductivity}
  &\sigma_{\text{el}} = \frac{m_D^2}{3\gamma}, \nonumber\\
  &\gamma = \frac{2g^2T}{4\pi}\left[ \ln\frac{m_D}{\gamma} + 3.041 + \mathcal{O}\left(\tfrac{1}{\ln(1/g)}\right) \right] \ ,
\end{align}
where $m_D^2$ is the Debye mass. In the SM $m_D^2 = (11/6)g^2T^2$ leading to the numerical value $\sigma_{\text{el}} = 0.91895 T$ (see \cite{Bodeker:2025ahg} for a recently computed QCD correction).
In principle, the BSM physics would effect $\sigma_{\textrm{el}}$ through modifying the matching relation for the Debye mass, as well as hard thermal loops used to derive Eq.~\eqref{eq:color-conductivity}, c.f.~\cite{Moore:1998zk,Moore:2000mx}. Such effects can be expected to have minor importance, and for simplicity we do not consider them further.

For details of update algorithms to measure statistical and dynamical parts of the sphaleron rate, we refer previous works \cite{Moore:1998swa,Moore:2000jw,DOnofrio:2012phz,Gould:2022ran}. Our goal is to determine the sphaleron rate in the entire $(x,y)$-plane below the critical $y_c(x)$ curve, i.e. in the Higgs phase. We directly jump into presenting our results.

\section{Results for the sphaleron rate}
\label{sec:results-lattice}

\begin{figure}[t]
  \includegraphics[width=1.0\columnwidth]{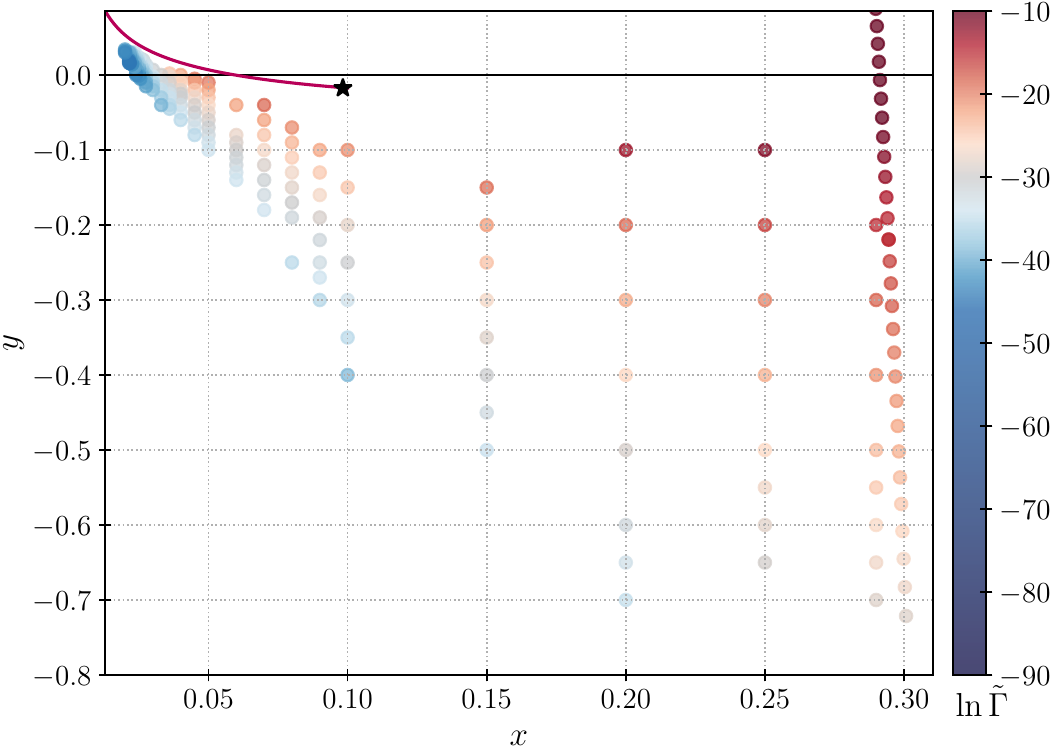}
  \caption{ The sphaleron rate for all the different $x$ and $y$ values we computed. These are for the smallest lattice spacing used $ag_3^2 = 0.25$ with volume $V=32^3a^3$. For each point the rate was computed from \eqref{sph_rate_lattice} with $\sim 5\times 10 ^5$ samples for the statistical part and $\sim 4000$ trajectories for the dynamical part.
   The data can be found in \cite{zenodo_data}. }
  \label{fig:rate_data}
  \vspace{-0mm}
\end{figure}
\begin{figure}[t]
  \includegraphics[width=1.0\columnwidth]{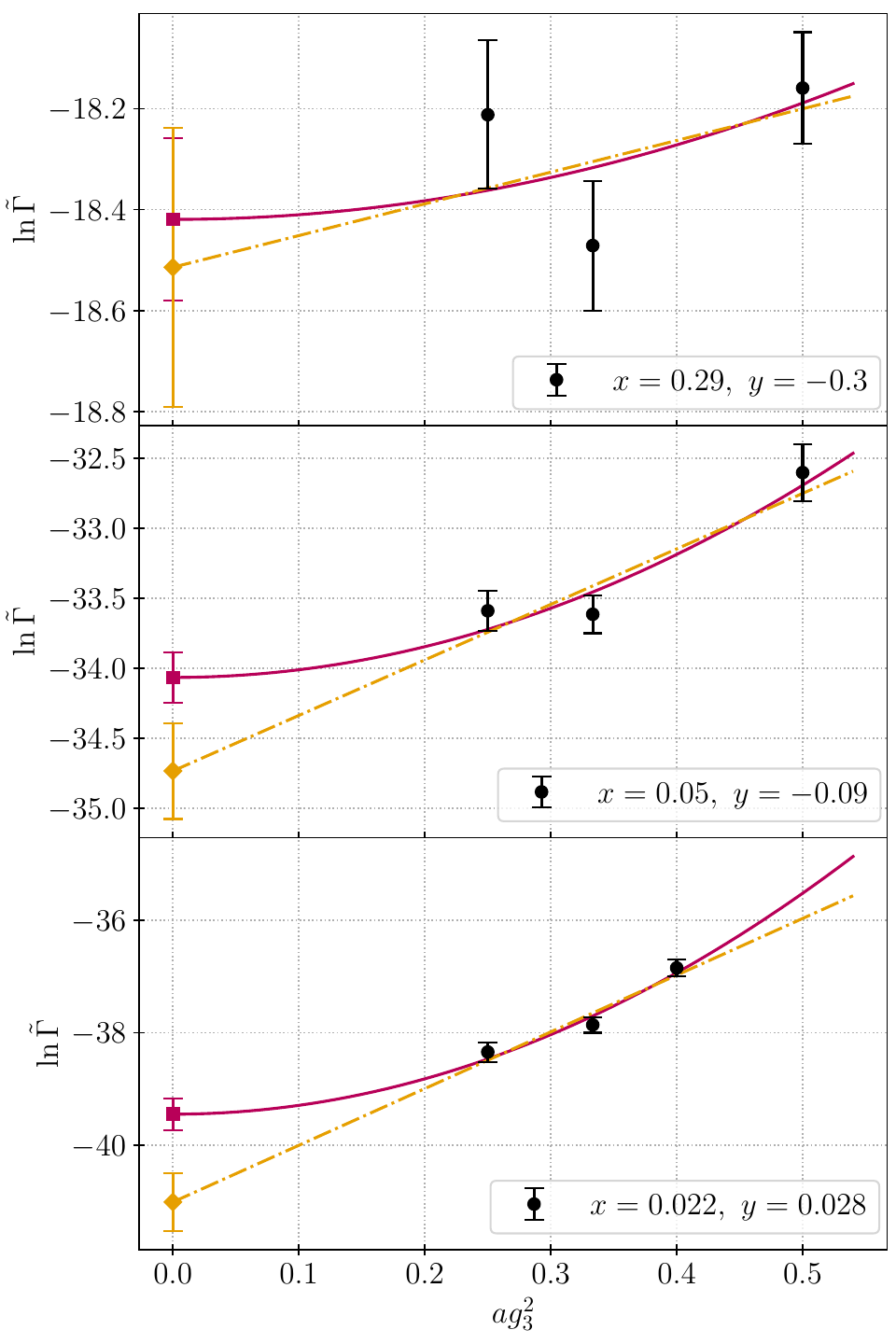}
  \caption{ The sphaleron rate for large and small $x$ values with different lattice spacings with fixed volume. We show both a linear $1+a$ and quadratic $1+a^2$ fits to the data points extrapolated to $a\to 0$. For large $x$ there is no noticeable lattice spacing dependence for the sizes simulated. For small $x$ there is noticeable dependence on the lattice spacing. For small $x$ the quadratic fits are used to estimate the error from not taking the continuum limit.}
  \label{fig:climits}
  \vspace{-0mm}
\end{figure}
\begin{figure*}[t]
  \begin{centering}
    \includegraphics[width=0.666\columnwidth]{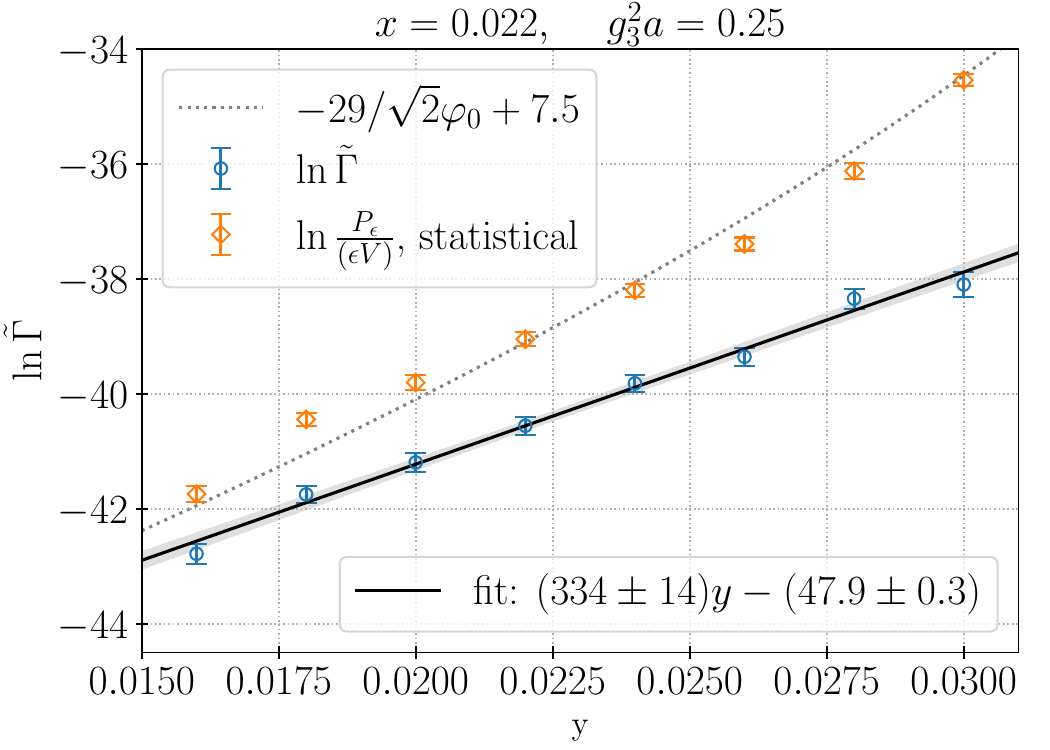}
    \includegraphics[width=0.666\columnwidth]{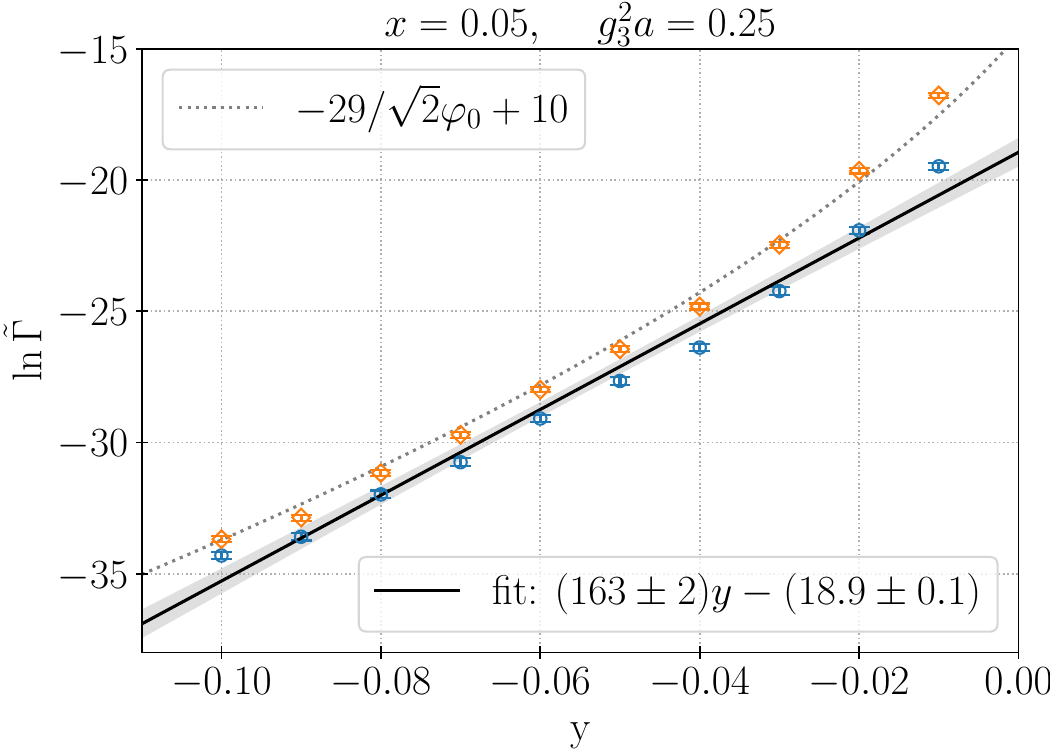}
    \includegraphics[width=0.666\columnwidth]{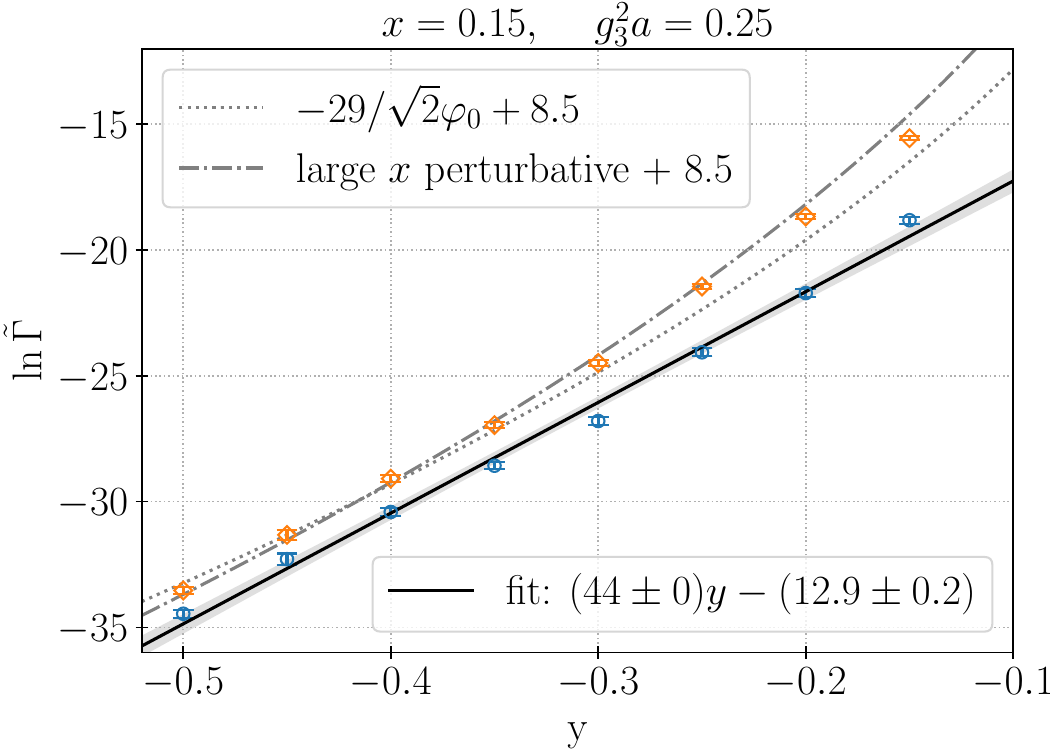}
  \end{centering}
  \caption{ 
Sphaleron rate at three different $x$ values as a function of $y$. The blue data points are the full rate and the orange points show only the statistical part as defined in \eqref{sph_rate_lattice}. The solid black lines are the fits to the full rate $\ln\tilde{\Gamma}$. The gray dotted curves are the estimates from a semi-classical computation without a dynamical factor (see Eq.~\eqref{rate_semiclassical}), shifted by an order $O(10)$ number inserted by hand, as explained in the main body. 
On the right most panel the dot-dashed line is the semi-classical computation for large $x$ values (see Eq.~\eqref{eq:large-x-pert}).
}
  \label{fig:fixed_x_rates}
\end{figure*}

Using the methodology of the previous section, we have computed the sphaleron rate for different values of $x$ and $y$ corresponding to the Higgs phase. 
The points where we have done simulations can be seen in Fig.~\ref{fig:rate_data} and the data can be found in \cite{zenodo_data}.

As in previous works we find the lattice spacing dependence to be mild for large values of $x$. However, for small $x \lesssim 0.15$ there now appears noticeable lattice spacing dependence. This is due to the fact that the sphaleron rate is very sensitive to the value of the Higgs condensate and thus the location of the phase transition, i.e.\ the critical value $y_c$, which does depend noticeably on the lattice spacing \cite{DOnofrio:2015gop,Gould:2022ran}. Thus, when the transition is of first-order, or the crossover is very sharply defined, around the critical $y_c$ value the Higgs condensate changes rapidly making the sphaleron rate more sensitive to the location of $y_c$. Or other way around, deeper in the crossover region the Higgs condensate changes more slowly around the $y_c$ value and thus its location matters less. In the crossover case $y_c$ here refers to the pseudocritical point defined as the location of the peak in the Higgs fields susceptibility.

We do not attempt to obtain a full continuum limit due to it requiring considerable computational efforts. Instead, for the small $x \lesssim 0.1$, where the lattice spacing has an effect, we computed the rate for a few different lattice spacings $ag_3^2 = 0.25, 0.333, 0.4, 0.5$ to estimate the error from not taking the continuum limit and used the smallest lattice spacing $ag_3^2 = 0.25$ for computing the results. The lattice volume $V=32^3a^3$ was chosen such that the finite volume effects are negligible \cite{Moore:1999fs, DOnofrio:2012phz, Annala:2023jvr}. See Fig.~\ref{fig:climits} for the general behavior of the rate on the lattice spacing for small and large values of $x$.

In Fig.~\ref{fig:fixed_x_rates} we illustrate our results for the sphaleron rate at three different fixed values of $x$ as function of $y$ below the corresponding $y_c$ value.
The full result for the rate is shown in blue with error bars showing the statistical error, together with a fit in black with standard error estimates in gray. In addition, we plot the statistical part of the rate as defined in Eq.~\eqref{sph_rate_lattice} in orange.
We can observe that the dynamical contribution to Eq.~\eqref{sph_rate_lattice} is significant but clearly subleading in comparison with the statistical part. However, as noted previously, the exact value of the statistical part will depend on the amount of cooling used in the Chern-Simons number measurement.

For further comparison, for each case in Fig.~\ref{fig:fixed_x_rates} we have plotted a semi-classical result of \cite{Li:2025kyo} (for details, see Appendix~\ref{sec:pert}). Such semi-classical approximation is oblivious to the real-time dynamics (described by the dynamical part of Eq.~\eqref{sph_rate_lattice}), but interestingly matches fairly well the lattice results for the statistical part up to a $y$-independent constants, by which we have shifted the perturbative results manually. This allows us to speculate that the magnitude of next-to-leading order (NLO) perturbative corrections described by the fluctuation determinant (see discussion in Appendix~\ref{sec:pert} and Ref.~\cite{Li:2025kyo}) would presumably match the constants (up to even higher-order contributions in perturbation theory). Furthermore, for this to be the case, such corrections would bare only minor $y$-dependence, and hence depend mostly on $x$ alone. We interpret these findings as a further motivation to extend results of \cite{Li:2025kyo} in future to include consistent determination of these NLO effects in purely perturbative approaches (see also \cite{Burnier:2005hp}).       

In Fig.~\ref{fig:rate_contours} (left panel) we present our result for the sphaleron rate $\ln \tilde{\Gamma}$ in the $(x,y)$-plane from $x=0.02$ up to $x=0.3$. The colored contours are obtained by linearly interpolating the data points obtained from simulations.
Around $x\approx 0.3$ we show the SM mapping $\{ x(T), y(T) \}$ for temperature range $T\simeq130-167$ GeV. At $x \lesssim 0.1$ red curve shows the non-perturbative results of $y_c(x)$ and the blue curve shows the $y_p(x)$ curve \cite{Gould:2022ran}.

\section{Cosmology: The Sphaleron Freeze-out}
\label{sec:results-cosmo}

\begin{figure*}
  \begin{centering}
    \includegraphics[width=1.0\columnwidth]{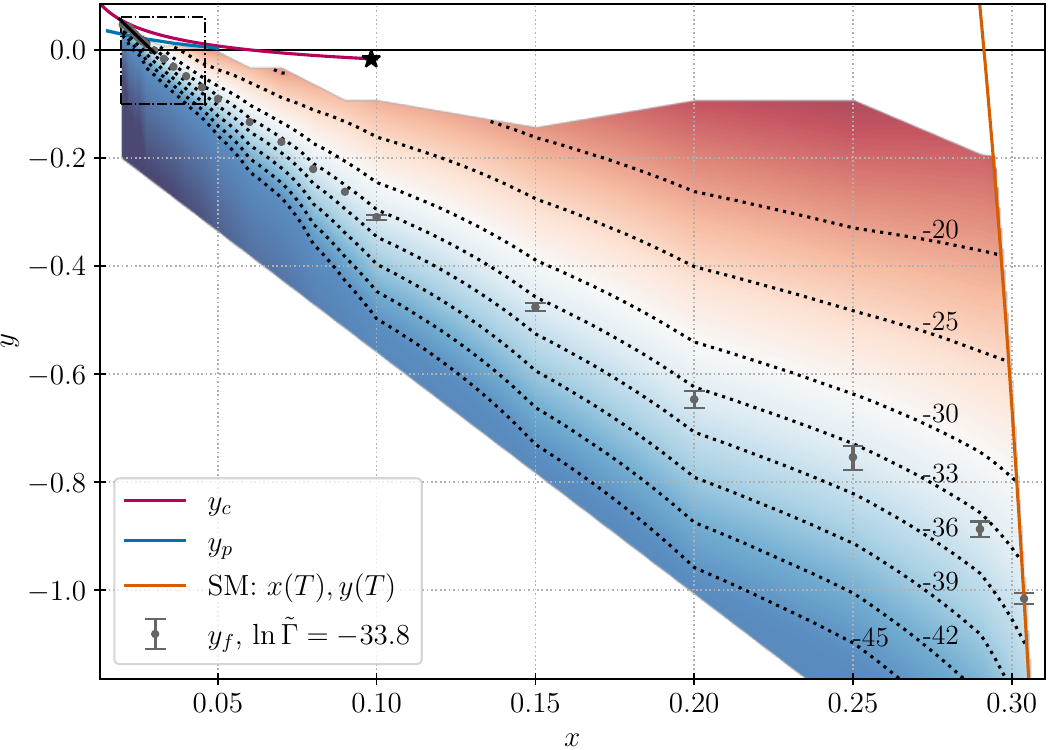}
    \includegraphics[width=1.0\columnwidth]{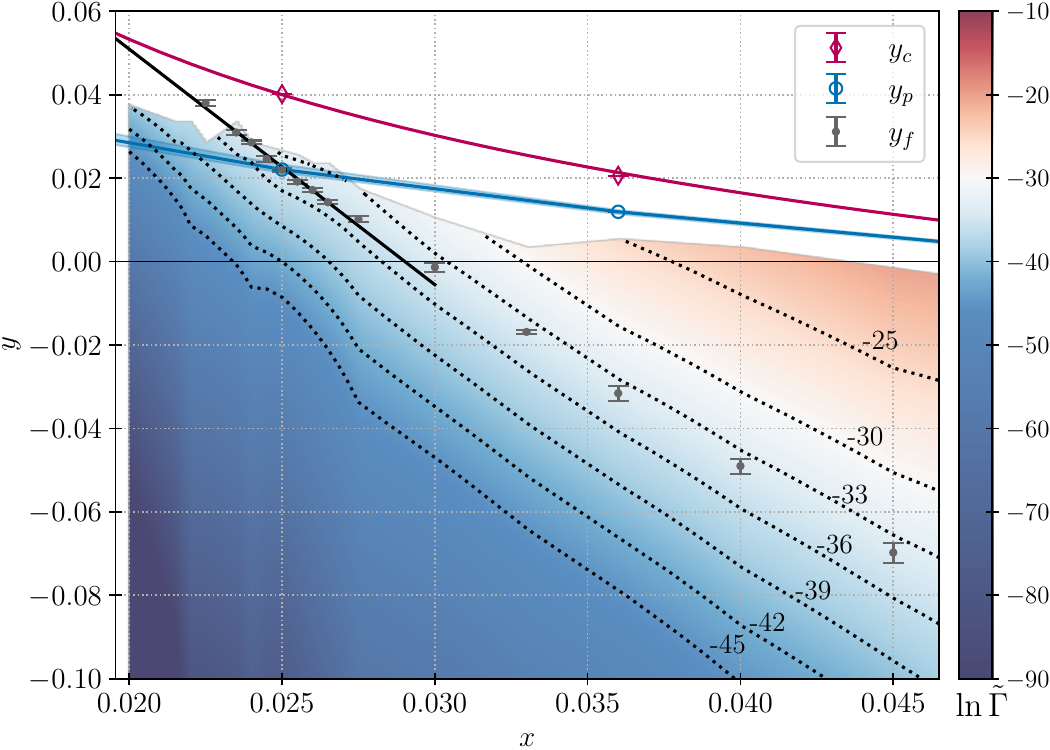}
  \end{centering}
  \caption{The sphaleron rate in the $(x,y)$ plane. The contour and black dotted constant rate lines are obtained from linearly interpolating the data obtained from simulations with $a g_3^2 = 0.25$, $V = 32^3 a^3$. The orange almost vertical line corresponds to the SM mapping for temperatures $130-167~$GeV. The gray dots with error bars are the sphaleron freeze-out $y$-values for fixed $x$, i.e.\ $y_f(x)$. The black solid line is a fit to small $x\lesssim 0.03$ values of $y_f$. The red and blue solid lines are the non-perturbative results for $y_c(x)$ and $y_p(x)$ respectively \cite{Gould:2022ran}. The star denotes the first-order phase transition end point $x_*=0.0983$ \cite{Gould:2022ran}, i.e.\ for larger values of
  $x$ the transition turns into a smooth crossover. On the left panel the dot-dashed box denotes the area that is zoomed in on the right panel which shows the relevant range for strong first-order phase transitions.}
  \label{fig:rate_contours}
\end{figure*}

Sphaleron freeze-out condition, at temperature $T_f$, can be written as \cite{Burnier:2005hp} 
\begin{align}
\label{eq:Tf-exact}
 N_\rho \frac{\Gamma(T_f)}{T^3_f} = H(T_f),
\end{align}
where $N_\rho \approx \frac{13}{4} \times 3$,%
\footnote{$N_\rho$ is a factor coming from the free energy of the system at fixed baryon number, see \cite{Burnier:2005hp, Khlebnikov:1996vj}.} with the Hubble parameter $H(T) = \sqrt{\frac{4 \pi^3 g_*}{45}} \frac{T^2}{m_{\text{pl}}}$, with $g_* = 106.75$ relativistic degrees of freedom%
\footnote{The value of $g_*$ can differ from the SM value used here in various BSM models. However, the effect is small even for $O(10)$ changes in $g_*$.} and the Planck mass $m_{\text{pl}} = 1.22 \times 10^{19}$ GeV. 
From Eq.~\eqref{eq:Tf-exact} we can solve $T_f$ for any dimensional reduction mapping exactly, but in analogy to definition of $y_p$ curve, we can get an approximate general result by noting that 
\begin{align} \label{rate_freeze_cond}
\ln \tilde{\Gamma}(x,y_f) + \ln \Big( \frac{ (g^2_3)^5}{\sigma_{\textrm{el}} T^4_f} \Big) = \ln \frac{H(T_f)}{N_\rho T_f}  \approx -38.3 \pm 0.5 ,
\end{align}
is approximately a constant for the relevant temperature range. For example, for $T=100$ to $250~$GeV right-hand side of \eqref{rate_freeze_cond} varies only from $-38.8$ to $-37.9$. 
Furthermore, we assume that $g^2_3$ and $\sigma_{\textrm{el}}$ are not significantly affected by the BSM physics.
This allows us to invert this condition, to find a curve $y_f(x)$.

For a fixed value of $x$ we are not able to efficiently simulate $y$ values that are very deep in the Higgs phase far from the critical $y_c$ value, and thus for larger $x$ values ($x \gtrsim 0.1$) we cannot reach the point of the freeze-out condition. However, the rate in the Higgs phase is well approximated by a pure exponential, and thus we can perform a linear fit to $\ln \tilde{\Gamma}(y)$ to extrapolate the rate deeper to the Higgs phase to find the $y_f$ freeze-out value (similarly what was done in the SM case \cite{D'Onofrio:2014kta, Annala:2023jvr}), see e.g.\ Fig.~\ref{fig:fixed_x_rates}. Regardless, for the interesting region of first-order transitions with small $x \lesssim 0.1$ we do not need to extrapolate to deeper in the Higgs phase. 

The curve of sphaleron freeze-out $y_f(x)$ is shown as the black line in Fig.~\ref{fig:rate_contours}, which turns out to be almost linear in the small $x<0.03$ regime.  
In the right panel of Fig.~\ref{fig:rate_contours} we zoom into the top left corner, relevant for strong first-order phase transitions. For very small values of $x$, the freeze-out line is above the percolation line $y_p(x)$. Because the percolation temperature  is close to the temperature where the bubbles nucleate, this means that the sphaleron freeze-out occurs immediately inside the growing bubbles.
\begin{figure}[t]
  \includegraphics[width=1.0\columnwidth]{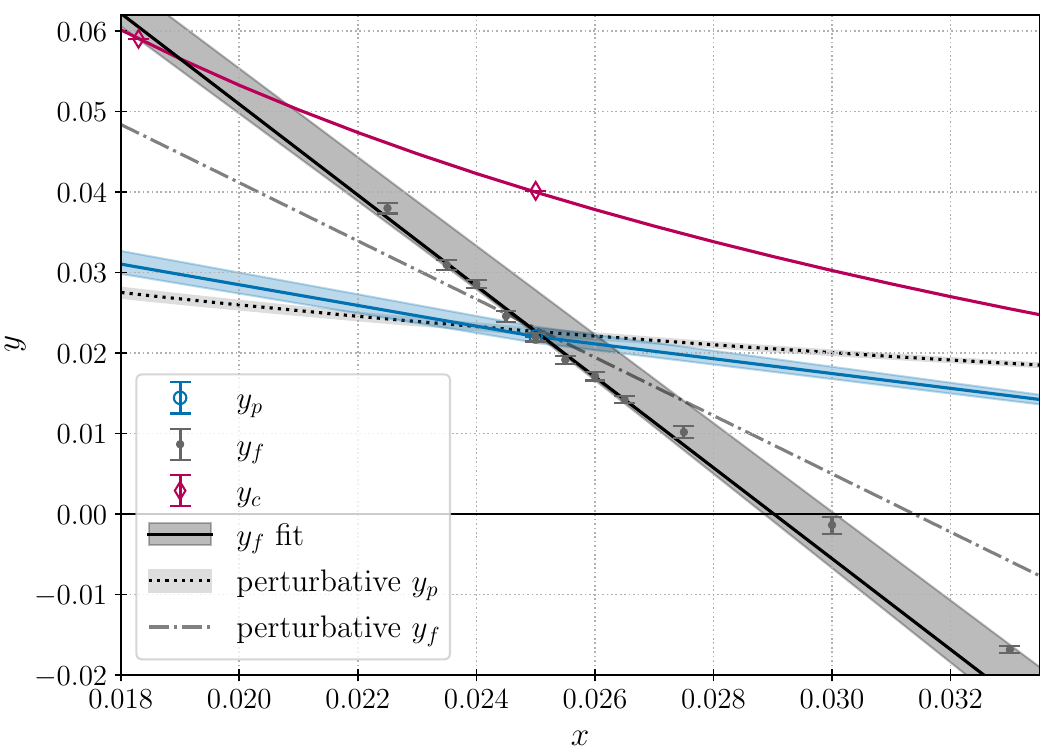}
  \caption{ 
Zoomed in plot of Fig.~\ref{fig:rate_contours} where the sphaleron freeze-out $y_f$ and the nucleation temperature $y_p$ curves intersect at $0.0251^{+0.0012}_{-0.0002}$. The error bounds for the blue $y_p$ curve are from varying the cosmology condition $\pm 5$. The black solid line is a linear fit to the obtained $y_f(x)$ values for $x < 0.03$ and the gray error bound is a combination of the fit error and the estimated continuum limit error. The linear fit for small $x$ is $y_f = (-5.65\pm 0.14)x + (0.164\pm 0.004)$.
}
  \label{fig:yc_yp_yf}
  \vspace{-0mm}
\end{figure}

This is further illustrated in Fig.~\ref{fig:yc_yp_yf}, and we find the intersection point of $y_p$ and $y_f$ curves at $ \bar{x} \equiv 0.0251^{+0.0012}_{-0.0002}$, where we have added errors from varying the cosmology condition and added continuum limit error estimates. 
Hence, the main result we find is (assuming weak $T$-dependence of $x$ in the dimensional reduction mapping) that all transitions with
\begin{align}
\label{eq:BNPC}
x_c \lesssim \bar{x} \approx 0.025,
\end{align} 
prevent the sphaleron washout inside the bubbles of the Higgs phase, and hence $x_c \lesssim \bar{x}$ guarantees baryon number preservation after strong first-order phase transition. We point out, that this condition is more strict compared to conditions $x_c \lesssim 0.04$ reported in the seminal work \cite{Kajantie:1995kf} as well as $x_c \lesssim 0.037$ obtained in \cite{Moore:1998swa}. This indicates, that for viable baryogenesis the electroweak phase transition has to be somewhat stronger than previously anticipated. 

Perturbative results for $y_p$ and $y_f$ are shown as black dotted and gray dot-dashed curves, respectively, and they also intersect at $x \approx 0.025$, as found in \cite{Li:2025kyo}. Given that the slope of the perturbative $y_f$ curve is slightly different from our non-perturbative result, this almost perfect agreement of the intersection point of $y_f$ and $y_p$ curves between these two results seems accidental. Nevertheless, this comparison demonstrates that the perturbative approach of \cite{Li:2025kyo} works fairly well.

In the next section, we give a general template with concrete examples how our result can be applied to electroweak phase transitions in Standard Model scalar extensions.

\section{Beyond the Standard Model phase transitions}
\label{sec:BSM}

In the SM, there is no phase transition due to relatively large value of the Higgs boson mass, resulting to a too large value for the self-coupling $\lambda$ to allow $x_c \lesssim 0.1$. This story can be altered, if new BSM physics can reduce the thermal Higgs self-coupling $\lambda_3$, and hence $x_c$ at and around the critical temperature.
Concretely, this can be achieved by introducing a new scalar field $S$ that couples to the Higgs scalar via a portal interaction
\begin{align}
\label{eq:Z2-portal}
a_2 \phi^\dagger \phi S^2,
\end{align} 
with portal coupling $a_2$, where both fields $\phi$ and $S$ are those of a full parent theory at zero temperature. To describe a general strategy to study phase structure of the model at high temperatures, we first remain agnostic about the concrete representation of $S$ under the SU(2) symmetry. First, we will assume a portal interaction quartic in the fields, and afterwards we discuss a cubic portal that leads to a mixing between the scalars.

In addition to Eq.~\eqref{eq:Z2-portal}, augmenting the SM with the new $S$ field introduces two other parameters: the mass (squared) $\mu^2_S$ and quartic self-interaction $b_4$ of $S$.
Often, it is convenient to trade the mass squared parameter to the physical pole mass of the $S$ field, or masses, in case of multiple fields. If we assume that there is only one such field (for instance in the case when $S$ is a one-component singlet) 
and that at the zero temperature minimum at leading order in perturbation theory $\langle \phi  \rangle \equiv v_0 \sim 246~$GeV and $\langle S \rangle = 0$ (which correspond to the typical electroweak minimum): we have a simple relation
$\mu^2_S = M^2_S - \frac12 a_2 v^2_0$ in terms of the pole mass $M_S$
at tree-level in perturbation theory. We note that the expectation value of the Higgs is not a physical quantity and hereby treat it only in the context of a tree-level analysis, while we return to discuss a more careful treatment below.

A general strategy to locate regions of first-order phase transitions in $(M_S,a_2,b_4)$-space, using non-perturbative information from the lattice simulations, proceeds as follows:
\begin{itemize}
\item[(i)] Identify a boundary between phenomenologically unviable region where the electroweak minimum is not the global minimum at zero temperature, and the region where the global minimum is the electroweak one.
\item[(ii)] Identify a boundary $\mu^2_S = 0$: In the region $\mu^2_S > 0$ the scalar $S$ in the 3D EFT cannot become light to undergo a transition of its own to have a non-zero expectation value at some high temperature range, while in the region $\mu^2_S < 0$ such a multistep phase transition is possible. 
\item[(iii)] At large $M_S$ and small $a_2$, the scalar $S$ decouples from the SM thermodynamics, and akin to SM there is no phase transition but a crossover. 
\item[(iv)] In-between regions (ii) and (iii), the $S$ field within the 3D EFT is generally heavy compared to the Higgs field at and around the critical temperature $T_c$, allowing to integrate it out. In such cases, the $S$ field modifies the $\{x(T),y(T)\}$-mapping. If at $T_c$ $x_c \lesssim 0.1$, then there is a first-order phase transition, otherwise a crossover (excluding the case $x_c < 0$ which indicates that some perturbative corrections are missing, potentially higher dimensional marginal operators).
As found in Sec.~\ref{sec:results-cosmo}, if $x_c \lesssim 0.025$ the transition is strong enough for viable EW baryogenesis. 
\end{itemize} 

This general strategy has been previously discussed or (at least partly) applied in \cite{Kajantie:1995kf,Cline:1996cr,Losada:1996ju,Laine:1996ms,Moore:1998swa,Brauner:2016fla,Andersen:2017ika, Gorda:2018hvi, Niemi:2018asa, Gould:2019qek,Friedrich:2022cak, Ramsey-Musolf:2024ykk,Li:2025kyo}. 

Before turning to concrete examples, let us discuss why in particular this approach is powerful. First, identifying (i) and (ii) requires nothing but zero temperature physics, meaning that these steps are free from complications arising at high temperature, such as accounting for thermal resummations that arise due to thermal screening and Bose-enhancement. On the other hand, the 3D EFT at high temperatures is exactly the device to account for such resummations consistently, relevant in (iii) and (iv).

The point (iii) is at first glance self-evident, but carries an insightful observation: at large $M_S$ and small $a_2$ the thermal mass of the $S$ field is large compared to thermal portal coupling, and hence it is certainly reliable to integrate out the $S$ field in this regime. This leads to an accurate, only slightly modified $\{x(T),y(T)\}$-mapping compared to the pure SM, as well as negligible marginal operators often dropped from the EFT. The striking point about (iv) is that any single-step first-order transition has to be located in-between regimes (i) and (iii), and a non-perturbative probe for this is the value of $x_c$, in-between (ii) and (iii). 

Hence, by applying the above strategy one can extract reliable partly non-perturbative information about the phase structure of any model without resorting to more conventional (and purely perturbative) approaches, such as direct numerical minimisation of a thermal effective potential which is numerically more expensive compared to working with the  $\{x(T),y(T)\}$-mapping, and unreliable for weak transitions unable to find a boundary between first-order transitions and crossovers. On the other hand, one still has to ensure that the perturbative part of the analysis, the dimensional reduction mapping, is reliable including the omission of marginal operators.%
\footnote{
For recent studies where marginal operators cannot be neglected, see \cite{Laine:2018lgj,Chala:2024xll,Bernardo:2025vkz,Chala:2025aiz}.
}
We discuss this aspect further below.  

Furthermore, once the boundary between crossovers and first-order transitions has been reliably located by the strategy outlined above, one can apply full thermal effective potential at two-loop order without resorting to integrating out the extra fields in parts of parameter space that admit strong transitions. This perturbative approach has been demonstrated to be in fair agreement with lattice simulations that \textit{include active} new scalar fields in the EFT, see \cite{Laine:2012jy,Kainulainen:2019kyp,Gould:2023ovu,Niemi:2024axp}. The benefit of such combined approach is, that it is numerically a lot more feasible to limit the use of a two-loop thermal effective potential to parts of parameter space that can already be anticipated to admit strong transitions, instead of scanning through the vast free parameter space by brute force.   

The strategy outlined above is general for all one-step phase transitions irrespective of a representation of $S$ or multiple such fields, and corresponding dimensional reduction mappings can be obtained in automated fashion \cite{Ekstedt:2022bff}. Before moving on, we emphasize that this strategy cannot be used to study multistep phase transitions with more complicated phase transition pattern, see e.g.~\cite{PhysRevD.9.3357,Land:1992sm,Hammerschmitt:1994fn,Patel:2012pi,Blinov:2015sna,Niemi:2020hto,Gould:2023ovu}, yet such transitions are located somewhere in-between boundaries (i) and (ii) that can themselves be found, as already emphasized, simply by zero temperature physics.%
\footnote{
Possible existence of multistep transitions in this region is intuitive in the following sense: in the region where the electroweak vacuum is not global, the system has to transition from the confinement phase to the $S$-phase at some high temperature, and only in a vicinity of this region it is possible that system transitions first to the $S$-phase at some temperature, and subsequently to the electroweak phase at some lower temperature. Where exactly this happens in-between boundaries (i) and (ii), requires to study thermal EFT with new scalars actively included \cite{Niemi:2020hto,Gould:2023ovu}. 
}

Next, we provide some concrete flavor, by assuming that the $S$-field is a singlet under SU(2), and has $N$ real components possessing an O(N)-symmetry. Akin to \cite{Tenkanen:2022tly}, we assume that the singlet vacuum expectation value at zero temperature vanishes. For studies where the new scalar is a doublet or triplet under SU(2), see \cite{Andersen:2017ika, Gorda:2018hvi, Niemi:2018asa,Li:2025kyo}. 
New terms in the scalar potential are 
\begin{align}
\label{eq:Z2-portal-singlet}
\frac12 \mu^2_S S^2 + \frac14 b_4 S^4 + a_2 \phi^\dagger \phi S^2 \ .
\end{align}  
As we have already discussed, we can relate the mass squared parameter to physical mass by $\mu^2_S = M^2_S - \frac12 a_2 v^2_0$. By shifting the fields as $\phi \rightarrow \phi + v/\sqrt{2}$ and $S \rightarrow S + s$ by real background fields, the requirement that the electroweak minimum of the tree-level potential, i.e.\ the minimum to $v$-direction, can be written as 
\begin{align}
-\frac{\mu^2}{4 \lambda} < -\frac{\mu^2_S}{4 b_4},
\end{align} 
required for (i), where $\mu^2$ is the Higgs mass (squared) parameter. 

\begin{figure*}
  \begin{centering}
    \includegraphics[width=1.333\columnwidth]{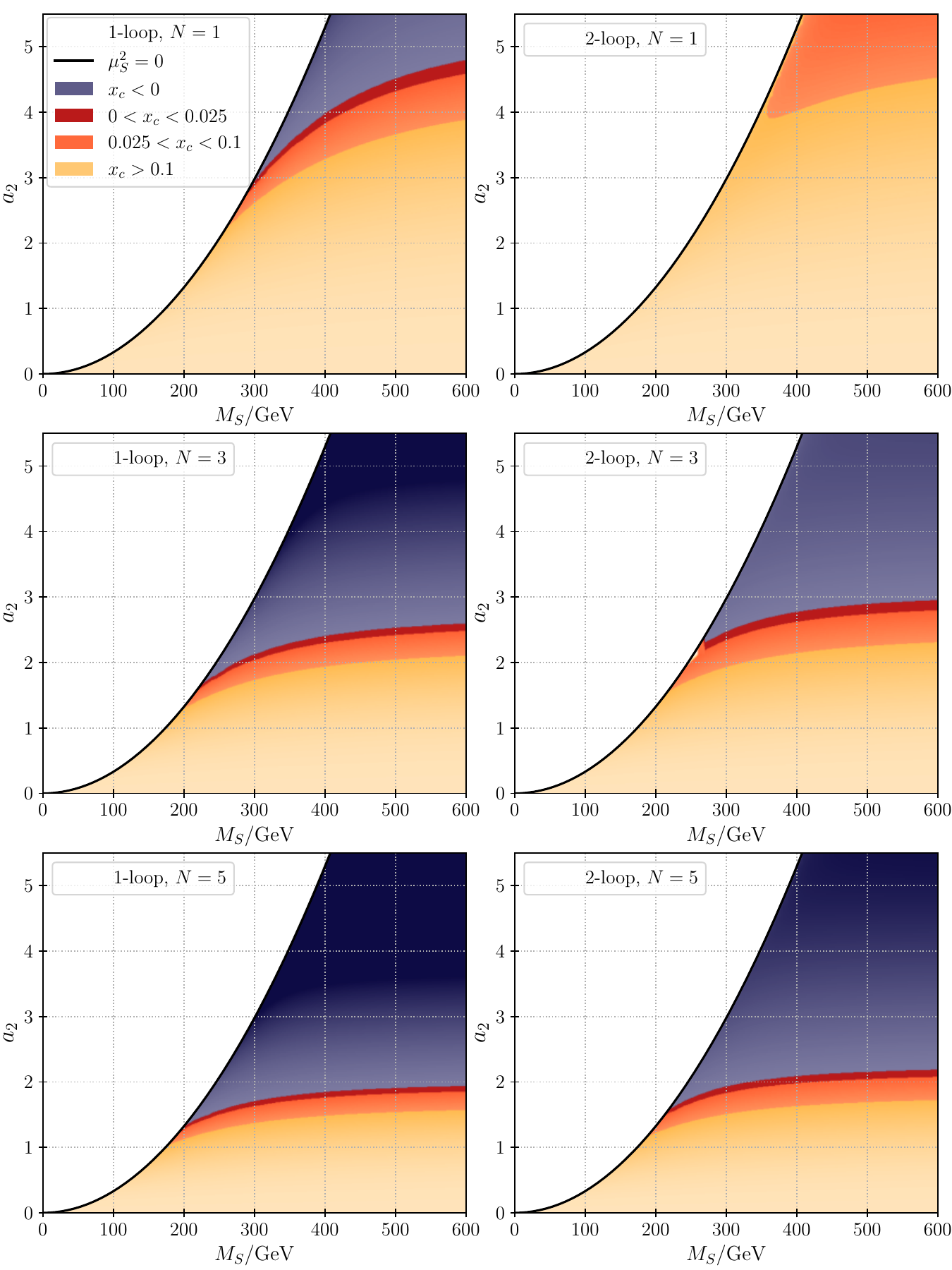}
  \end{centering}
  \vspace*{-0.2cm}
  \caption{
Regions of crossover in yellow ($x_c > 0.1$), weak first-order transitions in orange ($0.025 < x_c < 0.1$), strong first-order transitions in red ($0 < x_c < 0.025$) and regime where marginal operators or other higher-order effects are non-negligible in blue ($x_c < 0$) in $(M_S,a_2)$-plane with fixed $b_4 = 1$. In first column, soft contributions from the singlet are computed at one-loop, and in the second column at two loops, while different rows use $N=1,3,5$, in descending order. 
}
\label{fig:Z2-sym-example}
\end{figure*}

For constructing the EFT, we assume that singlet mass parameter is suppressed as $\mu^2_S \ll (\pi T)^2$ compared to the scale of non-zero Matsubara modes (referred later as the ``hard'' scale), and once the hard scale is integrated out the singlet thermal mass at one-loop reads \cite{Tenkanen:2022tly}
\begin{align}
\mu^2_{S,3} = \mu^2_S + \frac{1}{12} T^2 \Big( 2 a_2 + (N+2) b_4 \Big).
\end{align} 
We assume this thermal mass to be heavy compared to the scale of the transition, i.e. $\mu^2_{S,3} \sim m^2_D \sim (gT)^2$, and hence integrate out the singlet (akin to temporal gauge field modes).%
\footnote{
If the singlet vacuum mass parameter is very heavy ($\mu^2_S \sim (\pi T)^2$), the singlet needs to be integrated out entirely (including the zero Matsubara mode) in the dimensional reduction, see \cite{Laine:2000kv,Brauner:2016fla}. We do not discuss this situation here for simplicity, c.f.~\cite{Niemi:2018asa}.
}
The scale of the transition is set by the Higgs thermal mass $m^2_3 = y(T) g^4_3 \sim \frac{1}{x} g^4 T^2$ around the critical temperature $y_c \sim 1/x$. For small $x \sim g/\pi$ one hence has $m^2_3 \sim (g^{\frac32} T / \sqrt{\pi} )^2 $, which is parametrically in-between the soft $(g T)$ and ultrasoft $( g^2T / \pi)$ scales \cite{Gynther:2005av,Ekstedt:2022zro,Gould:2023ovu}.

Consequently, Higgs thermal self-coupling at the scale of the transition reads
\begin{align}
\label{eq:lam3}
\lambda_3 = T \Big( \lambda(\Lambda) - \beta_\lambda L_b \Big) - \frac{N}{32\pi} \frac{a^2_{2,3}}{\sqrt{\mu^2_{S,3}}} + \frac{N}{4(4\pi)^2} \frac{a^3_{2,3}}{\mu^2_{S,3}},
\end{align} 
where $a_{2,3} = T \Big( a_2 - \beta_{a_2} L_b \Big)$, and we have included singlet contributions from the hard scale $\sim \pi T$ at one-loop, and from the soft scale $\sim g T$ at two-loop level.%
\footnote{Contributions from the scale $g T$ are enhanced, and reaching the same formal accuracy as for the hard scale contributions requires a higher loop order calculation \cite{generic}.
}
Here $L_b \propto \ln(\Lambda/T)$ in the hard scale part, and $\beta$ denotes one-loop beta functions describing running with respect to the (parent theory) renormalisation scale $\Lambda$.  

The singlet contributions in Eq.~\eqref{eq:lam3} manifest in \textit{three} distinct ways: 1) soft scale corrections within the EFT (last two terms), 2) hard scale contributions proportional to $L_b$ and described by the running of $\lambda$; note that $\beta_{\lambda}$ includes a term proportional to $a^2_2$, see \cite{Schicho:2021gca,Niemi:2021qvp}, and 3) the value of $\lambda$ at the initial scale (often identified as the Z-boson mass) which receives corrections from the singlet via one-loop vacuum renormalisation, c.f.~\cite{Niemi:2021qvp}. As discussed in \cite{Niemi:2024vzw} (see also \cite{Moore:1995jv}), for accurate numerical analysis all these three effects are important and need to be taken in account. However, for our discussion here, we focus only on the soft scale corrections.   

At one-loop, the soft scale correction in Eq.~\eqref{eq:lam3} is described by the second last term, and crucially, contributes with a minus sign: this term reduces $\lambda_3$ and is exactly what is desired to strengthen the transition. On the other hand, the last term in Eq.~\eqref{eq:lam3} that arises at two loops from the soft scale, dilutes this desired reduction \cite{Ekstedt:2024etx,Niemi:2024vzw}, and turns out to be crucially important to account for, as we see below. We observe that in order to reduce $\lambda_3$ so that $x_c \lesssim 0.025$ these two terms need to be in delicate balance with the first term in Eq.~\eqref{eq:lam3}, and for this we observe below that having $N>1$ helps considerably. We point out that the soft scale perturbative expansion in Eq.~\eqref{eq:lam3} is an expansion in $a_{2,3}/(\pi \sqrt{\mu^2_{S,3}})$, and it is this ratio that needs to be small enough for integrating out the singlet to be sensible. 

The thermal mass of the Higgs, is composed as
\begin{align}
\label{eq:mass3}
m^2_3 &= m^2_{3,\text{SM}} + \frac{N}{24} T^2 a_2(\Lambda) - \frac{N}{8\pi} a_{2,3} \sqrt{\mu^2_{S,3}} \nonumber \\
&+ \frac{N}{4(4\pi)^2} a_{2,3} \bigg( (N+2)b_{4,3} - a_{2,3} \Big( 1 + \ln\Big[ \frac{\Lambda_3}{2 \sqrt{\mu^2_{S,3}} } \Big] \Big) \bigg),   
\end{align}
where we isolated the SM part (see e.g.~\cite{generic, Niemi:2021qvp}), $b_{4,3} = T ( b_4 - \beta_{b_4} L_b)$ \cite{Niemi:2021qvp} and $\Lambda_3$ is the renormalisation scale of the 3D EFT. We note that here the second term describes hard scale corrections at one-loop, but we did not write down the corresponding two-loop contribution from the hard scale, see \cite{Niemi:2021qvp}. In practical calculations, it is crucial to include such two-loop contribution to cancel the running of the one-loop term \cite{Arnold:1992rz,Gould:2021oba}. In Eq.~\eqref{eq:mass3} the last two terms are soft scale corrections from the singlet, at one and two loops, respectively.     

We note, that for generic models the dimensional reduction matching relations (as well as required one-loop $\beta$-functions) can be obtained using {\tt DRalgo} code \cite{Ekstedt:2022bff,Fonseca:2020vke}. However, zero-temperature vacuum renormalisation procedure required for relating Lagrangian parameters of a model to physical quantities such as pole masses (see e.g.~\cite{generic,Laine:2017hdk,Niemi:2021qvp}) is not included. Likewise, the soft scale corrections to scalar couplings (c.f. Eq.~\eqref{eq:lam3}) is not (automatically) included in the current version.%
\footnote{
Here we concretely refer to {\tt DRalgo} version 1.3. However, two-loop soft corrections to the scalar coupling (Eq.~\eqref{eq:lam3}) can be obtained (manually) with a following trick: {\tt DRalgo} can be used to compute the two-loop effective potential $V^{\text{3D}}_{\text{eff}}(v_3)$ with respect to the Higgs background field $v_3$, which includes soft scale corrections by a new scalar $S$. Since the effective potential is the generator of $n$-point Green's functions $G_n$, one can expand it as series in $v_3$, identify and isolate contributions of $S$ and extract the required 4-point function from $\frac{1}{4} v^4_3 G_4$ for the two-loop matching \cite{generic}.  
}

The critical temperature $T_c$ can now be determined from the condition $y(T_c) = y(x_c(T_c))$. That is, from the condition that $\{x(T),y(T)\}$-mapping intersects with (non-perturbatively determined) $y_c(x)$ curve at $T_c$. However, since $y(T)$ is fairly rapidly changing with temperature, below we resort to a simplification $y(T_c) \approx 0$ and use this condition to estimate the $T_c$. This condition is independent of $x(T)$, and hence provides a slight simplification for our numerical illustration below, in which we do not include relevant two-loop thermal masses, running of parameters, nor one-loop vacuum renormalisation. However, for any serious analyses that accounts these effects there is no reason to resort to this simplification.   

As an illustration of the previous discussions we fix $b_4 = 1$ and solve $T_c$ and $x_c$ in parts of $(M_S,a_2)$-plane where $\mu^2_S > 0$ (and hence also the electroweak minimum is automatically global). We depict result in Fig.~\ref{fig:Z2-sym-example} which shows different regions of crossover ($x_c > 0.1$), weak first-order transitions ($0.025 < x_c < 0.1$), strong first-order transitions ($0 < x_c < 0.025$) and regions where marginal operators or other higher-order effects are non-negligible ($x_c < 0$). In the left column we show results including only one-loop corrections of the singlet from the soft scale, while the right column also includes the two-loop (soft scale) contributions. For each row we use $N=1,3,5$ in descending order.

In top row with $N=1$ we observe that integrating out the singlet at one-loop (left) results in a narrow band of strong transitions with $0 < x_c < 0.025$, yet inclusion of two-loop corrections (right) completely removes this region, and even in the top right corner with large $a_2$ and $M_S$, $x_c$ does not become small enough for strong transitions. 
This result indicates that for $N=1$ there exists no strong one-step phase transitions which can be described within our approach. Yet we emphasize that our analysis does not indicate that this model with $N=1$ could not admit strong two-step transitions in which a light singlet plays a dynamical role and cannot be integrated out.

On the other hand, second and third row of Fig.~\ref{fig:Z2-sym-example} indicate that, as $N$ is increased, two-loop corrections do not spoil conditions for strong transitions. Furthermore, as could be expected by increasing the number of degrees of freedom that are integrated out, the value of portal coupling required to trigger the transition becomes smaller. This could perhaps be taken as a motivation to study BSM models where larger number of heavy fields are used to trigger the phase transition, such as Two- or Three-Higgs Doublet Models \cite{Keus:2013hya}, or even more exotic models with higher dimensional representation electroweak multiplets \cite{AbdusSalam:2013eya,Chao:2018xwz,Du:2018eaw}. 

However, we emphasize that results of Fig.~\ref{fig:Z2-sym-example} are only illustrative,%
\footnote{In other words, we use this $N$-singlet extension as an illustrative toy model, rather than as a serious candidate for the BSM theory.}
and do not include several higher order corrections that are expected to be important, such as two-loop thermal masses, one-loop running of parameters nor corrections from one-loop vacuum renormalisation. Indeed, in \cite{Li:2025kyo} it has been concluded that when such corrections are included in the real-triplet extended SM \cite{Patel:2012pi, Niemi:2018asa, Niemi:2020hto,Gould:2023ovu} all strong one-step transitions found based on including only one-loop soft scale corrections from the triplet, vanish when two-loop effect are included (see also \cite{Niemi:2024vzw} for similar conclusions in the real singlet-extended SM).

\begin{figure*}
  \begin{centering}
    \includegraphics[width=2.0\columnwidth]{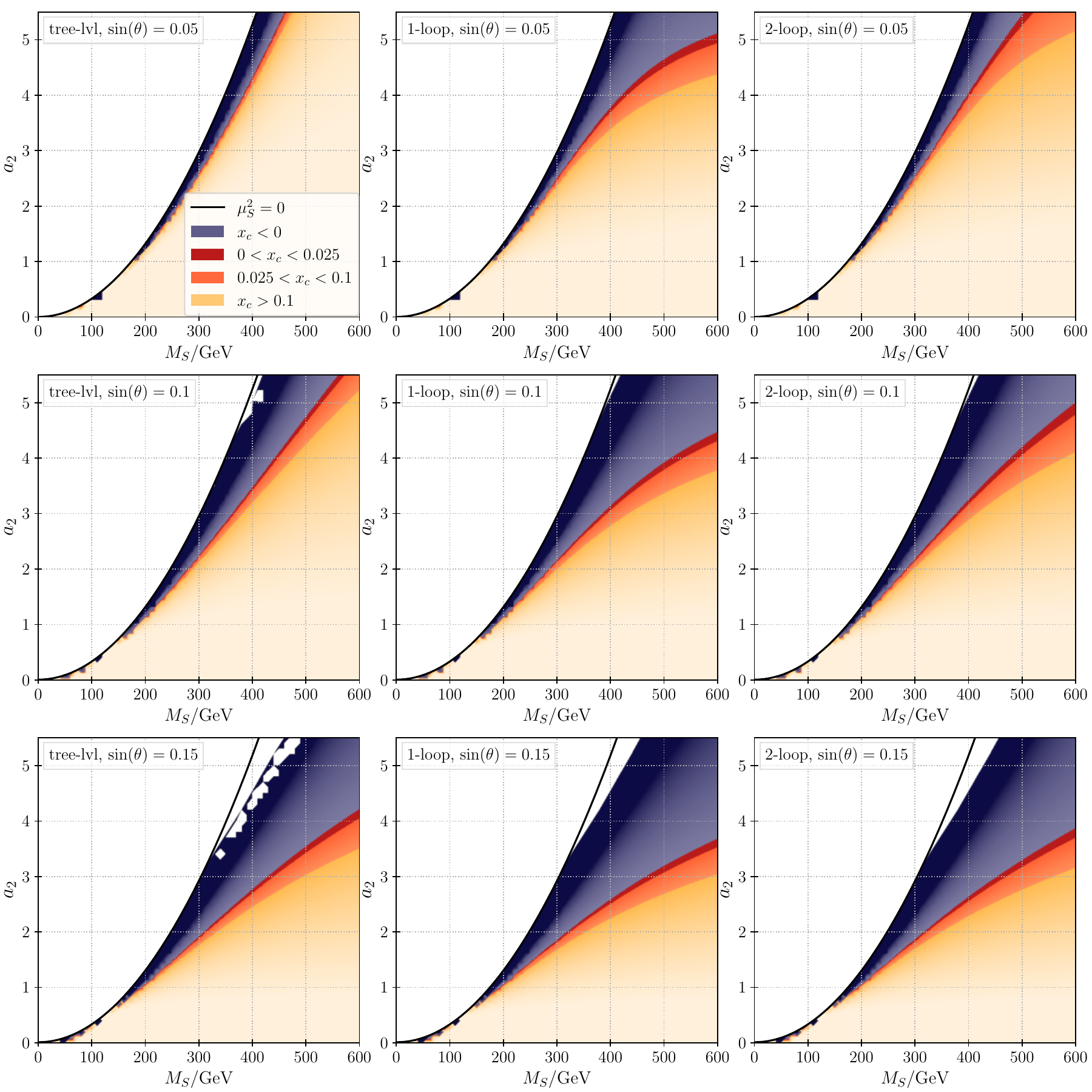}
  \end{centering}
  \caption{
Similar to Fig.~\ref{fig:Z2-sym-example}, but in non-$Z_2$ symmetric case with $N=1$ and with varying the mixing angle.
}
\label{fig:non-Z2-sym-example}
\end{figure*}

Next, we shall discuss a possibility that the singlet model includes non-$Z_2$ symmetric operators \cite{Choi:1993cv,Profumo:2007wc}, that is, we add the following terms to the scalar potential
\begin{align}
b_1 S + b_3 S^3 + a_1 S \phi^\dagger \phi.
\end{align}  
From now on for simplicity, we also assume that $N=1$.
Due to the presence of the cubic portal, two scalars mix with a mixing angle we denote by $\theta$. In terms of physical masses the Lagrangian parameters read (at tree-level)  
\begin{align}
\mu^2 &= -\frac14 \Big( M^2_h + M^2_S + (M^2_h - M^2_S) \cos(2\theta) \Big) \,, \nonumber \\
\mu^2_S &= \frac12 \Big( -a_2 v^2_0 +  M^2_h + M^2_S - (M^2_h - M^2_S) \cos(2\theta) \Big) \,, \nonumber \\
\lambda &=\frac{1}{4v^2_0} \Big( M^2_h + M^2_S + (M^2_h - M^2_S) \cos(2\theta) \Big) \,, \nonumber \\
a_1 &= -\frac{2}{v_0} (M^2_h - M^2_S) \sin \theta \cos\theta\,, \nonumber \\
\label{eq:tadpole}
b_1 &= -\frac14 a_1 v^2_0 \ .
\end{align}
Again, for any serious investigations these relations should be augmented with corrections from one-loop vacuum renormalisation, c.f.~\cite{Niemi:2021qvp}. From this reference, one can also find a physically meaningful recipe to replace the unphysical Higgs vacuum expectation value $v_0$ by physical observables (the Fermi constant related to muon lifetime and weak gauge boson pole masses). Above, the relation of Eq.~\eqref{eq:tadpole} tying together $b_1$ and $a_1$ follows from convention that singlet vacuum expectation value at zero temperature vanishes, which can be achieved by a coordinate transformation shifting the singlet field. 
We note that without $Z_2$-symmetry ($S \rightarrow -S$), the presence of mixing -- which we assume to be small $\sin(\theta) \ll 1$ in accord with phenomenology -- in fact slightly increases the Higgs self-interaction coupling $\lambda$.   
We also comment that, in presence of the mixing, the condition for the electroweak minimum to be global becomes significantly more complicated analytically. Nevertheless, it is straightforward to guarantee this condition numerically.    

Cubic couplings $b_3$ and $a_1$ have dimension of mass, and for dimensional reduction to be reliable these couplings have to be small enough compared to the temperature (for details see \cite{Niemi:2021qvp}). Since we keep the mixing angle small, in practice for $a_1$ this requires the pole mass $M_S$ to not be exceedingly large. Within the EFT, these operators are described by couplings $b_3 = \sqrt{T} \Big( b_3 - \beta_{b_3} L_b \Big)$ and $a_1 = \sqrt{T} \Big( a_1 - \beta_{a_1} L_b \Big)$ at one-loop \cite{Niemi:2021qvp} while the tadpole coupling at one-loop is given by
\begin{align}
b_{1,3} &= \frac{1}{\sqrt{T}} \Big( b_1 + \frac{1}{12} T^2 (b_3 + a_1) \Big)\ .
\end{align}  
Again, accurate analyses would require one to include also the two-loop correction, but here we omit it for simplicity. 

\begin{figure*}
  \begin{centering}
    \includegraphics[width=1.0\columnwidth]{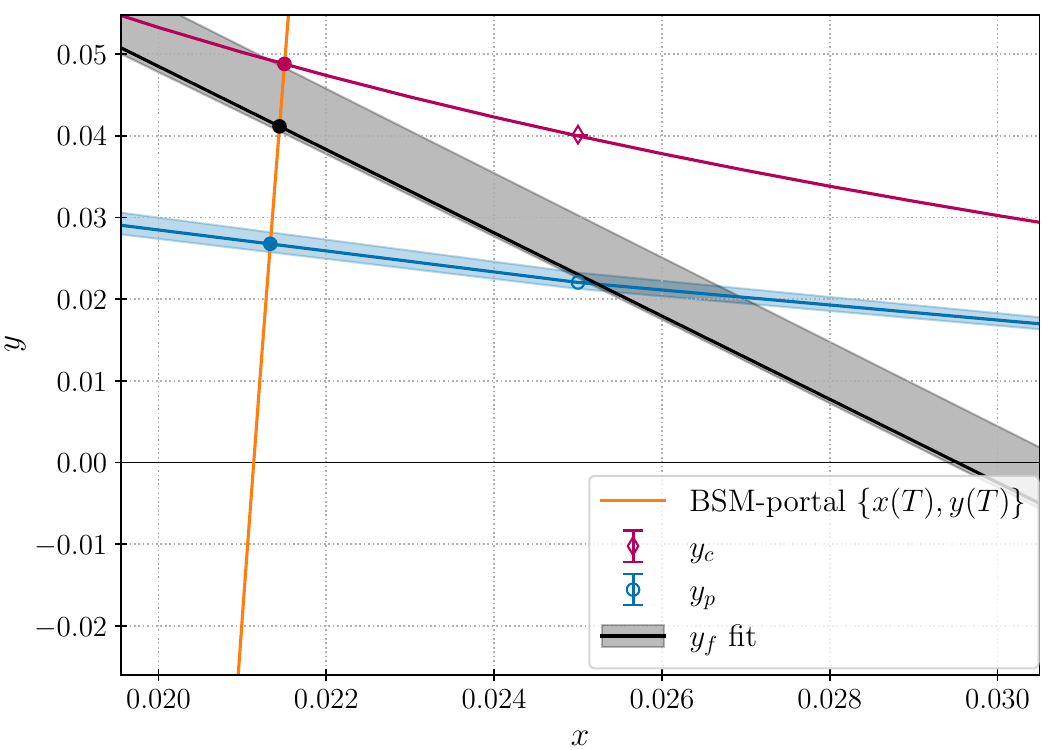}
    \includegraphics[width=1.0\columnwidth]{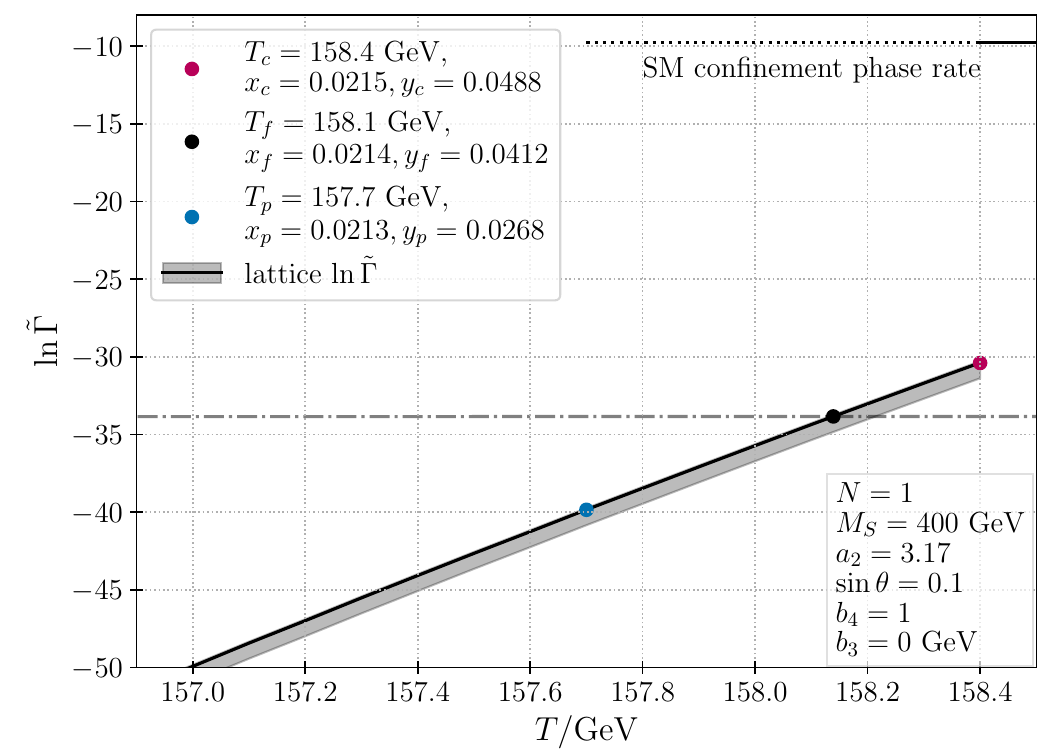}
  \end{centering}
  \caption{
On the left panel the orange curve shows the $\{x(T),y(T)\}$-mapping for a non-$Z_2$ symmetric $N=1$ scalar, with $M_S = 400~$GeV, $a_2 = 3.17$, $\sin\theta = 0.1$, $b_4=1$, $b_3 = 0$, and two-loop corrections; as in the middle row of Fig.~\ref{fig:non-Z2-sym-example}. The shown $\{x(T),y(T)\}$-mapping corresponds to temperature range of $155.9 - 158.6~$GeV. On the right panel we used the interpolated lattice data to get the sphaleron rate for the $\{x(T),y(T)\}$-mapping of the specific model, giving us the sphaleron rate as a function of temperature. The gray standard error bounds are again from the continuum limit error estimate. The horizontal dotted line shows the confinement phase sphaleron rate in the SM case as found in \cite{Annala:2023jvr}, here scaled to the dimensionless $\tilde{\Gamma}$. The gray dot-dashed line on the right plot shows the Hubble rate terms in \eqref{rate_freeze_cond} without approximating them as constants. }
  \label{fig:BSM_mapping}
\end{figure*}

When the singlet is integrated out at the soft scale within the EFT, already at tree-level, it modifies the Higgs thermal mass by a term \cite{Ekstedt:2024etx}
\begin{align}
- \frac{b_{1,3} a_{1,3}}{2 \mu^2_{S,3}},
\end{align}
and the Higgs thermal self-coupling by
\begin{align}
\label{eq:lam3-cubic}
- \frac{a^2_{1,3}}{8 \mu^2_{S,3}} + \frac{1}{4} b_{1,3} \Big( 2 \frac{a_{2,3} a_{1,3}}{\mu^4_{S,3}} - \frac{b_{3,3} a^2_{1,3}}{\mu^6_{S,3}} \Big)\ .
\end{align}
It turns out that these tree-level contributions dominate over the loop corrections involving cubic couplings \cite{Ekstedt:2024etx,Brauner:2016fla}, and hence for simplicity we do not consider such higher order corrections here. 
We observe that in particular the first term in Eq.~\eqref{eq:lam3-cubic} leads to efficient reduction of $x_c$.

Next, we turn to numerics and fix $b_4 = 1.0$ and $b_3 = 0$ (varying these numbers would keep our discussion the same qualitatively) and similar to Fig.~\ref{fig:Z2-sym-example} we plot different regions based on $x_c$ in the $(M_S,a_2)$-plane for three different values $\sin(\theta) = 0.05,0.1,0.15$. We observe that again the top right part of the $(M_S,a_2)$-plane admits strong first-order transitions, and for increasing mixing angles, such regions move towards smaller $a_2$, as could have been anticipated. 

We emphasize that transitions can be expected to be of first-order (with increasing strength for increasing $a_2$) also in the region where $x_c <0$. This expectation is supported by perturbation theory within another EFT that has negative quartic self-coupling, yet small and positive coefficient for a sextic marginal operator $(\phi^\dagger \phi)^3$ \cite{Croon:2020cgk, Camargo-Molina:2021zgz, Camargo-Molina:2024sde}. However, non-perturbative studies of such EFT -- which suffers from absence of super-renormalisability and hence exact lattice-continuum relations -- have yet to be performed.

In the non-$Z_2$ symmetric case a key difference to the previous $Z_2$-symmetric case is, that since the reduction of $x_c$ is dominated by the cubic portal coupling that acts already at tree-level within the EFT, the results can be expected to be significantly less sensitive to higher order corrections, and hence more reliable. This provides evidence that as far as strong electroweak phase transitions are considered, models with cubic portals to Higgs appear perhaps much better motivated compared to models with mere quartic portals. On the other hand, such cubic portals cannot be incorporated to wide classes of models; this includes models with extra doublets as such operators are forbidden by gauge symmetry, or e.g. the real triplet model, where the gauge symmetry allows for cubic operator $\phi^\dagger (S^b \sigma_b /2)  \phi$, but which is tightly constrained to be small by the electroweak $\rho$-parameter \cite{Patel:2012pi}.

In Fig.~\ref{fig:BSM_mapping} we illustrate results for a single representative parameter point (defined in the plot legend), that maps into $x_c < 0.025$ (left), and allows us to extract the sphaleron rate as function of temperature (right). We observe, that indeed the temperature dependence in $x(T)$ is weak in the range from $y_c$ to $y_f$ and $y_p$, and that $T_c$, $T_f$ and $T_p$ are all within 1 GeV around 158 GeV, indicating very minor supercooling down from $T_c$. This is a typical trend for these heavy (thermal) portal scalar scenarios.

In summary, in this section we described a general, powerful strategy to study and find strong first-order electroweak phase transitions using combination of perturbation theory (dimensional reduction and integrating out extra scalars)
and non-perturbative information from the lattice, in particular for the character of a transition and whether it is sufficiently strong to preserve the baryon asymmetry by shutting off the sphaleron transitions. We further illustrated how this procedure can be concretely realised in a simplified setting with an extra singlet.

Yet we emphasize, that in order to simplify the discussion, we did not consider all higher order perturbative effects that can be expected to be important, such as two-loop thermal masses, renormalisation group running of parameters nor corrections from one-loop vacuum renormalisation (c.f.~\cite{Niemi:2021qvp}). For any serious studies, we advocate that these effects should be included, as well as estimates of the validity of dimensional reduction that could be compromised when high-temperature expansions and expansions at the soft scale become unreliable when large couplings or extremely strong transitions are involved \cite{Laine:1999rv,Laine:2000kv,Laine:2017hdk}.

\section{Discussion}
\label{sec:conc}

In this article we have extended previous non-perturbative determinations of the Higgs phase sphaleron rate in the minimal SU(2) + Higgs 3D EFT to entire parameter region in $(x,y)$-plane (corresponding to thermal self-coupling and thermal mass) relevant for first-order electroweak phase transitions in BSM theories in which first-order phase transition is triggered by a new field with large enough thermal mass. In particular, we determined non-perturbatively a curve $y_f(x)$ that captures information when the sphaleron transitions freeze out in the Higgs phase. Our study hence complements previous non-perturbative studies for analogous curves $y_c(x)$ and $y_p(x)$ describing the critical and percolation temperatures, respectively. 

Combining this information allowed us to confirm that $y_p(x)$ and $y_f(x)$ curves intersect, and hence find a condition that sphalerons in the Higgs phase are shut off before the percolation occurs, and this happens when $x_c \lesssim 0.025$, assuming that $x(T)$ is only mildly changing with temperature in-between $y_c=y(T_c)$ and $y_p=y(T_p)$.%
\footnote{
Should this assumption be relaxed, then the more general condition to prevent the sphaleron washout is $x_f = x(T_f) < 0.025$ and hence $y_f > y_p$, i.e. $T_f > T_p$. 
}
Under this assumption, the condition $x_c \lesssim 0.025$ provides a necessary,%
\footnote{
Strictly speaking, the condition $x_c \lesssim 0.025$ guarantees that sphalerons cannot wash out any baryon asymmetry in the Higgs phase, while at slightly larger values of $x_c$ it could be still possible to survive from the double exponential washout, if the initial asymmetry generated in front of the bubble is huge compared to experimentally observed value. For details, see \cite{Li:2025kyo}.
}
yet not sufficient condition for viability of the electroweak baryogenesis. This condition is slightly stricter than previously obtained by analogous investigations \cite{Kajantie:1995kf,Moore:1998swa}, and hence our results indicate that to preserve the baryon asymmetry the electroweak phase transition has to be slightly stronger than previously thought. 

So far we have assumed that the temperature in the Higgs phase inside the bubbles does not reheat significantly above $T_p$. However, relaxing this assumption and allowing the temperature to grow back all the way up to $T_c$ would provide an even stronger bound on phase transition strength. Using our lattice data we find $x_c \gtrsim 0.019^{+0.001}_{-0.0005}$ (stronger than reported in \cite{Rubakov:1996vz}), yet we comment that to obtain a more accurate value for this bound would require more careful analysis of the continuum limit of the $y_f(x)$ curve.

We can also translate the above upper bound on $x_c$ to a lower bound on the  discontinuity of the Higgs condensate $\Delta \langle \phi^\dagger \phi \rangle_c$ \cite{Farakos:1994xh} defined as the difference of the condensate in the different phases at the critical temperature (or $y_c$). For the explicit definition of $\langle \phi^\dagger \phi \rangle$ see e.g.\ \cite{Annala:2023jvr} Eq.\ (14). Using $x_c=0.025$ results to a lower bound $v/T_c \equiv \sqrt{2 \Delta\langle \phi^\dagger \phi \rangle_c g^2_3}/T_c \gtrsim 1.33$ (by using the results from \cite{Gould:2022ran} to obtain the discontinuity).\footnote{Note, here we have written the condensate in units of $g_3^2$ which is what is naturally obtained from the lattice. In the abstract the condensate is directly in 4D units.}
Here we also used the SM value $g_3^2\simeq 0.4 T$.
By computing the (gauge-invariant) condensate in perturbation theory for $x=0.025$, we obtain $v/T_c \gtrsim 1.01$ at leading order, and $v/T_c \gtrsim 1.36$ at $\text{N}^4\text{LO}$ using results of \cite{Ekstedt:2022zro,Ekstedt:2024etx}, with latter providing a fair agreement with the lattice.

In particular, these conditions provide a non-perturbative, consistent alternative to widely circulated, heuristic condition $v_c/T_c \gtrsim 1$ (in Landau gauge) to prevent the washout, in terms of the Higgs field expectation value $v_c$ at the critical temperature, computed from the thermal effective potential (see e.g.~\cite{Basler:2016obg,Ahriche:2007jp}). 
Furthermore, we emphasize that our non-perturbative condition is universal within the underlying assumptions about the construction of the EFT at the infrared, namely that everything but the SU(2) gauge fields and the Higgs doublet can be integrated out. In addition, we assumed that effects of BSM fields to the dynamical part of the rate are negligible.

We also saw that for small values of $x \lesssim 0.15$ the sphaleron rate does depend noticeably on the lattice spacing, unlike for large values of $x$ as seen in previous studies. This is because the sphaleron rate is very sensitive to the value of the Higgs condensate. Thus, when the transition is very sharply defined or of first-order, in the vicinity of the transition the value of the condensate, and by extension the sphaleron rate, is sensitive to the exact location of the critical temperature or $y_c$ which does depend on the lattice spacing.

To give flavor how our findings can be used to study electroweak phase transitions in a class of BSM theories under these assumptions, we also reviewed a general procedure and exemplified it in terms of a simplified setting with singlet scalar portal. In particular, we demonstrated how the most simple quartic portal is sensitive to higher-order perturbative corrections, yet argued that this feature can be mitigated if such portal comprises of multiple fields or fields that are in higher dimensional representations of electroweak multiplets. Furthermore, we showcased the attractive nature of a possible cubic portal to trigger strong electroweak phase transition, a feature that can be realised in e.g.\ the non-$Z_2$ symmetric singlet-extended SM, for which phenomenological signatures are actively predicted and searched for in the present and future collider experiments, c.f. e.g.~\cite{Niemi:2024axp,Antusch:2025lpm} and references therein.   

Interestingly, a semi-classical approximation of Ref.~\cite{Li:2025kyo} agrees fairly well with our non-perturbative determination. 
This can be understood in terms of the mass scales of the problem:
the mass scale of the sphaleron in the Higgs phase is set by the gauge boson mass, and hence well above the mass scale of the transitioning scalar \cite{Ekstedt:2022zro,Lofgren:2023sep,Gould:2023ovu}, hinting that a perturbative description of the sphaleron is faster converging than that of the bubbles \cite{Gould:2021ccf,Hirvonen:2022jba,Li:2025kyo}. Nonetheless, to concretely confirm this, it would be highly motivating to revisit a computation of the sphaleron fluctuation determinant \cite{Carson:1989rf,Carson:1990jm,Baacke:1993aj} to go beyond the semi-classical approximation within the EFT (also c.f.~\cite{Burnier:2005hp}). Furthermore, given the success of semi-classical approximation, it would certainly be motivating to generalize the semi-classical approximation of Ref.~\cite{Li:2025kyo} to a class of multistep phase transitions, that involve multiple active scalar fields charged under SU(2) within the final EFT at the infrared. Reliable perturbative approach to such more complicated situations would offer immensely faster ways to analyse vast free parameter spaces, compared to non-perturbative simulations.

\section*{Acknowledgments}

We thank
Andreas Ekstedt,
Johan L{\"o}fgren,
Xu-Xiang Li, 
Lauri Niemi,
Michael Ramsey-Musolf,
Van Que Tran,
Yanda Wu
and
Guotao Xia 
for enlightening discussions.
This work has been funded by the European Union (ERC, CoCoS, 101142449), and the Research Council of Finland grant 354572.
We wish to acknowledge CSC -- IT Center for Science, Finland, for computational resources.
\appendix

\section{Temporal gauge field components}
\label{sec:temporal}

In the effective theory of Eq.~\eqref{eq:continuum_theory} temporal gauge field components $A^a_0$ have been integrated out perturbatively, and their effect is encoded in the parameters of the EFT \cite{generic}. 
This procedure relies upon an approximation that the Debye mass $m_D \approx N_{m_D} g^2 T$ of $A^a_0$ fields dominates over the field-dependent mass generated by the Higgs mechanism. 
For very strong transitions (at very small $x$), that involve large jump in the Higgs condensate, this perturbative approximation becomes unreliable.
Technically, the previous criteria amounts to $m^2_D \gg h_3 v^2_3$, where $h_3 \approx \frac{1}{4} g^2 T$ is the coupling of $A^a_0$ to Higgs field, and $v_3$ is the scalar background field (used to generate correlation functions in the EFT matching). For small values of $x$, in the Higgs phase we can estimate the size of the background field by the leading order perturbative result $v_3 \approx \frac{g_3}{8\pi x}$ at the critical temperature \cite{Ekstedt:2022zro}, and hence we arrive to a lower bound 
\begin{align}
x \gg \frac{g}{16\pi \sqrt{N_{m_D}}} \approx 0.0093,
\end{align} 
for the SM values $g^2 \approx 0.4$ and $N_{m_D} \approx \frac{11}{6}$.
While this lower bound is indeed smaller than our condition $\bar{x} \approx 0.025$ for the baryon number preservation in Eq.~\eqref{eq:BNPC}, in the future it could be interesting to revisit our simulations without integrating out the temporal gauge field components (c.f.~\cite{Kajantie:1993ag,Farakos:1994kj,Kajantie:1995kf}), in order to scrutinize the accuracy of our present results. We also note, that in BSM theories where a new portal scalar is charged under SU(2), it would increase the value of $N_{m_D}$ (c.f.~\cite{Gorda:2018hvi,Niemi:2018asa}) and hence integrating out the temporal gauge fields becomes more reliable.

Additionally, as pointed out in \cite{Li:2025kyo}, previous simulations \cite{DOnofrio:2012phz,DOnofrio:2014rug,Annala:2023jvr} have not included the perturbative effect of temporal gluon fields to the Higgs thermal mass (see also \cite{Brauner:2016fla, Schicho:2021gca}). This effect scales as $\sim g^2_Y g^3_s/\pi^3$ and appears formally at higher order than NLO in the dimensional reduction \cite{generic}. Yet, it has been demonstrated to have a small, but not entirely negligible, effect on thermodynamics and the sphaleron rate \cite{Li:2025kyo}, due to relatively large numerical values of the top quark Yukawa coupling $g_Y$ and the strong coupling $g_s$.

\section{Semi-classical calculation in terms of Higgs condensate in perturbation theory}
\label{sec:pert}

In Ref.~\cite{Li:2025kyo} a semi-classical calculation%
\footnote{
In this approach, a sphaleron field configuration is found at classical level within the 3D EFT. This corresponds to a leading order approximation in perturbation theory, which would receive corrections from a fluctuation determinant computed for the sphaleron field configuration.  
}
of the sphaleron rate has been formulated for $x\gg 0.1$ as well as for $x<0.1$. In the latter case
\begin{align}
\label{rate_semiclassical}
\Gamma^{\text{LO}}_{\text{sph}} = T^4 e^{-S_{\text{sph}}(x,y)},
\end{align}
where the sphaleron action has a simple approximation $S_{\text{sph}}(x,y) \approx \frac{29}{\sqrt{2}} \varphi_0(x,y)$ in terms of the Higgs phase minimum of the leading order effective potential \cite{Moore:2000jw,Ekstedt:2022zro}
\begin{align}
V_{\text{LO}} = g^6_3 \Big( \frac{1}{2} y \varphi^2 + \frac{1}{4} x \varphi^2 -  \frac{1}{16\pi} \varphi^3 \Big),
\end{align}
where the scalar background field $\varphi$ is scaled to be dimensionless. Extremal condition $\frac{d}{d\varphi} V_{\text{LO}}|_{\varphi=\varphi_0}$ leads to the expression
\begin{align}
\varphi_0 = \frac{3 + \sqrt{9-1024 \pi^2 x y}}{32 \pi x},
\end{align} 
for the Higgs phase minimum. Note that our normalisation for the background field here differs by a factor $1/\sqrt{2}$ from the expression in \cite{Li:2025kyo}. 
On the lattice, the expectation value of the square of the Higgs field is described by the scalar condensate $\langle \phi^\dagger \phi \rangle$. This condensate can be computed in perturbation theory as \cite{Farakos:1994xh}
\begin{align}
\langle \phi^\dagger \phi \rangle \equiv \frac{d}{dy} V_{\text{eff}},
\end{align} 
and at leading order
\begin{align}
\label{eq:correspondence}
\varphi_0 = \sqrt{2\langle \phi^\dagger \phi \rangle}.
\end{align} 
At higher orders within the 3D EFT perturbation theory \cite{Farakos:1994kx}, it has been shown that this condensate can be computed with remarkable agreement with the lattice result, for strong phase transitions at small $x$ \cite{Ekstedt:2022zro,Ekstedt:2024etx}.

The correspondence in Eq.~\eqref{eq:correspondence} might tempt one to approximate the semi-classical sphaleron action in terms of the Higgs condensate, rather than the background field at the minimum as 
\begin{align}
\label{eq:rate-pert}
S_{\text{sph}}(x,y) \approx \frac{29}{\sqrt{2}} \sqrt{2\langle \phi^\dagger \phi \rangle},
\end{align}
with a following motive: 
two-loop corrections to the Higgs condensate are known to be large, 
and hence it would be desirable to add such corrections to the semi-classical formulation of \cite{Li:2025kyo}.
For this, we follow \cite{Ekstedt:2022zro,Ekstedt:2024etx} and use the three-loop effective potential, expanded around its leading order minimum;  this is gauge-invariant in accord with \cite{Nielsen:1975fs,Fukuda:1975di}.
Next, we compute the condensate by taking the y-derivative of the resulting potential, and we obtain the condensate as function of $x$ and $y$ at all five perturbative orders up to and including $\text{N}^4\text{LO}$ (c.f.~\cite{Ekstedt:2024etx}, in which similar results are reported in a strict expansion around leading order $y_c$). This results in somewhat lengthy analytic expressions, which we do not write down explicitly here. 

\begin{figure}[t]
  \includegraphics[width=1.0\columnwidth]{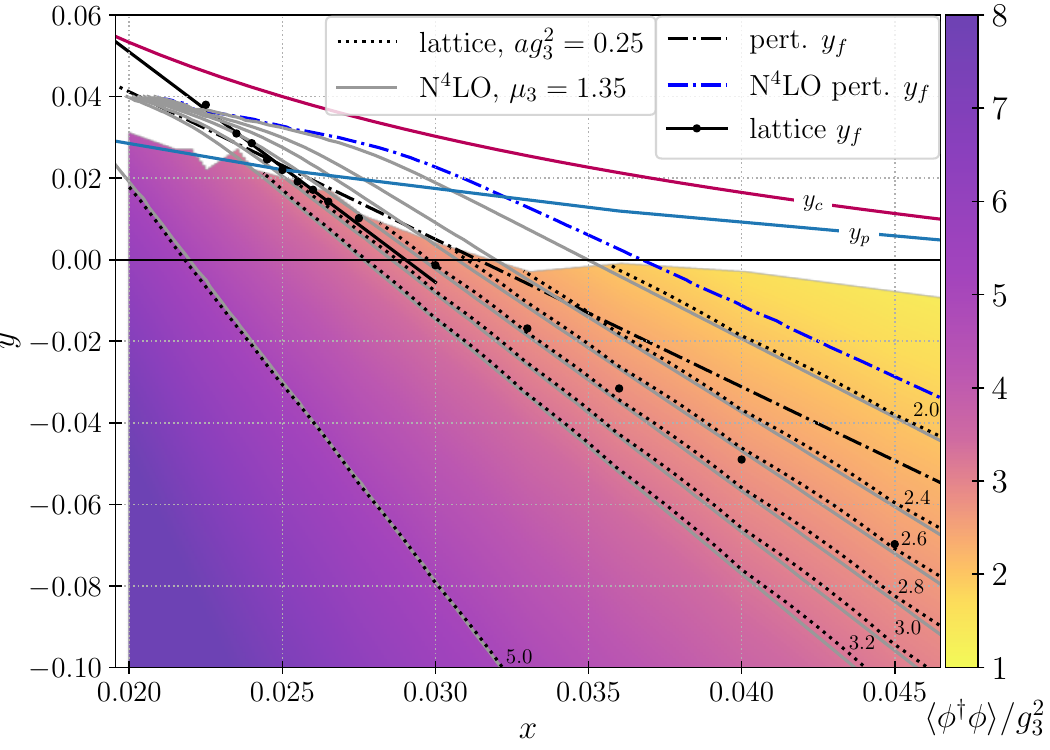}
  \caption{ 
Dimensionless condensate $\langle \phi^\dagger \phi \rangle/g^2_3$ in the $(x,y)$-plane, from lattice (dotted black) versus $\text{N}^4\text{LO}$ perturbation theory (solid gray), together with $y_f$-line from the lattice (solid black) and in perturbation theory with LO (dot-dashed black) and $\text{N}^4\text{LO}$ condensate (dot-dashed blue).   
}
  \label{fig:phi2_contours}
  \vspace{-0mm}
\end{figure}

We compare these perturbative results against the lattice results for the condensate in Fig.~\ref{fig:phi2_contours}, 
yet have to immediately stress the following: the condensate itself is not (3D) renormalisation scale invariant, while the difference of the condensate between two phases is \cite{Farakos:1994xh}. In this sense, our following observations are only to be taken as tentative. In perturbation theory, we fix the 3D renormalisation scale as $\Lambda_3 = 1.35 g^2_3$, and this choice matches the lattice result in Fig.~\ref{fig:phi2_contours} remarkably well (and point out, that perturbation theory converges rapidly at small $x$, with already NLO (two-loop) being close to lattice result, albeit we do not illustrate this). 

Using the action of Eq.~\eqref{eq:rate-pert} for the semi-classical approximation, we additionally show the results for $y_f$-curves using the condensate at LO and $\text{N}^4\text{LO}$.    
We observe that higher order corrections to the scalar condensate do not drastically change the result for perturbative $y_f$,
yet brings the perturbative result farther away from the lattice result. We stress, however, that simply adding such higher order corrections to the condensate as we have discussed, does not provide renormalisation scale invariant result for the sphaleron rate. 

What this exercise illustrates, on the other hand, is that 
in perturbation theory the sphaleron rate is not crucially sensitive to the higher order corrections that modify the Higgs expectation value (or the condensate) at two loops and beyond.
This conclusion can be viewed in terms of the physical picture provided by \cite{Gould:2021oba}: the mass scale of the transitioning scalar is in-between perturbative soft and non-perturbative ultrasoft scales, and major perturbative corrections to the thermodynamics arise from the soft scale through heavy gauge field in the Higgs phase. The mass scale of the sphalerons in the Higgs phase, on the other hand, is set by the gauge boson mass, and hence sphaleron dynamics are not similarly sensitive to large soft scale corrections, as sphalerons in the Higgs phase live at the soft scale themselves. For this reason, already the semi-classical approximation agrees reasonably well with the full non-perturbative result from the lattice. 
This leads us to speculate that leading corrections to the semi-classical picture, that are encapsulated by the one-loop fluctuation determinant computed around the sphaleron background \cite{Carson:1989rf,Carson:1990jm,Baacke:1993aj,Moore:1995jv}%
\footnote{Ref.~\cite{Moore:1995jv} discusses a computation of the fermionic fluctuation determinant analytically in terms of higher-dimensional, marginal operators within the EFT, and their effect to the sphaleron rate. Recent works \cite{Chala:2024xll,Bernardo:2025vkz,Chala:2025aiz,Chala:2025oul} study marginal operator effects on thermodynamics and bubble dynamics of phase transitions, yet do not consider the sphaleron rate.}
are responsible for the major of the remaining difference between the lattice and perturbation theory. This speculation is further supported by perturbative results reported in \cite{Burnier:2005hp} and related discussion in \cite{Li:2025kyo}, that our discussion here complements.  

Finally, for completeness we collect here the perturbative result of \cite{Li:2025kyo} for the sphaleron rate in the semi-classical approximation for large $x \gg 0.1$ used in Fig.~\ref{fig:fixed_x_rates}. That is, the classical action can be approximated as
\begin{align}
\label{eq:large-x-pert}
S_{\text{sph}}(x,y) \approx \sqrt{\frac{-y}{2x}} \Big( A + B x^C \ln(Dx) \Big), 
\end{align}
where the overall factor is simply the expectation value of the tree-level potential, and the remainder is a fit with coefficients $A=26.12$, $B=-2.145$, $C=0.4237$ and $D=0.00717$. This concludes our detour to perturbation theory.

\newpage 

\bibliography{main}

\end{document}